\documentclass[lettersize,journal]{IEEEtran}

\usepackage{amsmath,amssymb,amsfonts}
\usepackage{graphicx}
\usepackage{textcomp}
\usepackage{xcolor}

\usepackage{algpseudocode}
\usepackage{algorithm}
\usepackage{pifont}
\usepackage{eqparbox}
\usepackage{caption}
\usepackage{subcaption}
\usepackage{physics}
\usepackage{lipsum}
\usepackage{xspace} 
\usepackage{multirow}
\usepackage{tabularray}
\usepackage{makecell}

\usepackage{mathtools}
\DeclarePairedDelimiter\ceil{\lceil}{\rceil}

\def \ScalingActionIdx              {\alpha}
\def \ScalingActionSet              {A}

\def \CAPA          {\textsc{cap}}
\def \CPU           {\textsc{cpu}}
\def \COMP          {\textsc{comp}}

\def \LinkIdx           {\ell}
\def \LinkSet           {L}
\def \LinkCapacity      {\textsc{cap}_{\LinkIdx_{\SliceIdx}}}
\def \lstmfsd           {\textit{LSTM-FSD}}
\def \lpkpi             {LP_{\textsc{kpi}} }

\def \MaxCAPVNFSlice     {\textsc{cap}^{\max}_{\SliceIdx \ResourceIdx}}
\def \MinCAPVNFSlice     {\textsc{cap}^{\min}_{\SliceIdx \ResourceIdx}}
\def \MonitorTrafficTime        {t'}

\def \HistoricalData                {H_{\textsc{data}}}
\def \HistoricalDataNormalized      {H^{Norm}_{\textsc{data}}}

\def \NumberOfSampling      {\kappa}
\def \NumberOfVnf           {\gamma}

\def \PMIdx      {\textsc{pm}}
\def \PMSet      {\text{PMset}}
\def \PSA        {\textsc{PCLANSA}} 
\def \SetPhyMachine         {\textsc{PM}s}
\def \PhyMachineIdx         {\textsc{PM}}
\def \PHY                   {\textsc{PHY}}

\def \ratio                                  {\rho}
\def \ratioOverProvision                     {\ratio^{\textsc{op}}} 
\def \ratioScaling                           {\ratio^{\textsc{s}}} 
\def \ratioResUsage                          {\ratio^{\textsc{ru}}} 
\def \ratioValidateScaling                   {\ratio^{\textsc{d}}} 

\def \ResourceIdx                            {r}
\def \ResourceSet                            {R}
\def \RAM                                    {\textsc{ram}}
\def \RequireComputeResourcePerThroughput    {\text{Req}_{\SliceIdx}}

\def \SliceIdx      {s}
\def \SliceSet      {S}
\def \STO           {\textsc{sto}}
\def \SliceConfig   {s^{\ResourceIdx, \LinkIdx}}
\def \SliceResourceConfig   {s^{\ResourceIdx}}

\def \TimeWindow                {t}
\def \Throughput                {\textsc{th}}
\def \PredictedThroughput       {\widehat{\Throughput}}
\def \TrafficModelAccuary       {\varepsilon}
\def \TrafficErrorRate          {E_{\MonitorTrafficTime}}
\def \TrafficBoundary           {B^{\PredictedThroughput}_{\SliceIdx}}
\def \TargetSliceKPIs           {KPI^{\SliceIdx}}
\def \totalPhyCap               {\CAPA_{\PMSet}} 
\def \totalLinkCap              {L} 

\def \LinkUtilization                   {u^{\LinkIdx}}
\def \ResourceUtilizationVector         {U}
\def \ResourceUtilizationIdx            {\ResourceUtilizationVector\ResourceIdx}

\def \VNFInstance                   {\textsc{vnf}\textit{i}}
\def \VNFIdx                {\textsc{$v$}}
\def \VNFSet                {\text{VNFset}\textit{i}}

\newcommand{\cmark}{\ding{51}}%

\newcolumntype{C}[1]{>{\centering\arraybackslash}p{#1}} 

\newif\ifhighlighting
\highlightingtrue

\begin{document}
\title{Proactive Service Assurance in 5G and B5G Networks: A Closed-Loop Algorithm for End-to-End Network Slices}

\author{
    \IEEEauthorblockN{
        Nguyen Phuc Tran,
        Oscar Delgado (\IEEEmembership{Member, IEEE}),
        Brigitte Jaumard (\IEEEmembership{Senior Member, IEEE})
    }
    
    \IEEEauthorblockN{
        Computer Science and Software Engineering, Concordia University, Montr\'eal (Qu\'ebec), Canada\\
        Email for correspondence: brigitte.jaumard@concordia.ca
    }
}


\maketitle


\begin{abstract}
Ensuring the highest levels of performance and reliability for customized services in fifth-generation (5G) and beyond (B5G) networks requires the automation of resource management within network slices.
In this paper, we propose \PSA, a proactive closed-loop algorithm that dynamically allocates and scales resources to meet the demands of diverse applications in real time for an end-to-end (E2E) network slice.
In our experiment, \PSA \xspace was evaluated to ensure that each virtualized network function is allocated the precise resources it requires, thereby maximizing efficiency and minimizing waste.
This goal is achieved through the intelligent scaling of virtualized network functions.
The benefits of \PSA \xspace have been demonstrated across various network slice types, including eMBB, mMTC, uRLLC, and VoIP.
This finding indicates the potential for substantial gains in resource utilization and cost savings, with the possibility of reducing over-provisioning by up to 54.85\%.
\end{abstract}

\begin{IEEEkeywords}
5G Network Slice, Resource Allocation, Virtualized Network Functions (VNFs), Quality of Service (QoS), Proactive Resource Management, Closed-Loop Control, Dynamic Scaling, Machine Learning in 5G and B5G networks.
\end{IEEEkeywords}

\section{Introduction}

\IEEEPARstart{T}he increasing demand for diverse, high-performance applications in 5G and B5G networks necessitates efficient resource allocation and service assurance within network slices.
While network slicing offers customized service delivery, ensuring that each network slice meets its performance requirements (e.g., low latency for uRLLC, high throughput for eMBB) while minimizing resource consumption presents a significant challenge. 
Existing closed-loop service assurance mechanisms often react to performance degradations, leading to potential Service Level Agreements (SLAs) violations and inefficient resource utilization, as highlighted in recent studies \cite{9741386,kaur2021intelligent}.
To meet the evolving requirements of both network operators and end-users, these challenges must be effectively addressed.
In particular, B5G networks will demand greater load adaptability and scalability to support the rapid growth of 5G and B5G applications.
As a result, telecommunication networks have undergone substantial transformations in recent years to deliver higher speeds, enhanced reliability, and more responsive data transmission. 
Within this context, machine learning plays a pivotal role by enabling proactive network behaviour through real-time prediction, anomaly detection, and intelligent decision-making.

\IEEEpubidadjcol

In network slicing, the infrastructure must demonstrate the capacity to dynamically allocate resources in accordance with the service requirements of each network slice. These services encompass a broad spectrum of quality of service (QoS) needs, as detailed in the reference \cite{foukas2017network}.
In order to meet the QoS requirements, resource allocation is the common process of allocating specific resources, such as central processing units (CPUs), memory, and storage, to virtual network functions (VNF) instances.
This allocation can be executed manually or automatically.
Manual allocation requires a greater investment of time and is more susceptible to errors, but it provides better control over resource usage.
Automated allocation can be more efficient but may not always allocate resources optimally.
Thus, the VNF auto-scaling process entails a delicate balancing act between network actions and spare resources, with the objective of meeting QoS requirements while achieving cost savings, as articulated in the work of Rahman \textit{et al.} \cite{Rahman2018AutoScaling}.

\IEEEpubidadjcol

Network performance and resource utilization are often optimized with closed-loop control mechanisms in network slice \cite{3GPP2020Closed}.
The closed-loop algorithm is a feedback control system used in Service Assurance (SA) of communication networks to improve network performance and maintain service quality.
Closed-loop control mechanisms play a vital role in continuously monitoring performance and resource utilization, enabling real-time responses to satisfy the distinctive requirements of each network slice in functional domains such as the radio access network (RAN), transport network (TN), and core network (CN).
Achieving service assurance can involve modifying network configurations, better resource allocation, or improved management of traffic flows.
For instance, if a network slice needs more capacity, the control loop management and orchestration system can assign additional resources to the network slice in real-time without affecting the effectiveness of other network slices.

The employment of a closed-loop algorithm for 5G and B5G network slices yields numerous advantages, including the optimization of resources and network efficiency, as well as enhanced network performance through reduced latency and jitter.
Furthermore, network efficiency can be increased by mitigating congestion and service disruptions.
However, the development of a closed-loop algorithm also poses several challenges, as outlined in \cite{a2021_solutions}.
The network often has a high degree of complexity and is subject to various factors that can affect its operational efficiency.
Thus, numerous research activities are currently in progress to devise closed-loop algorithms for network slices in the 5G and B5G networks, organized by various entities such as academic institutions, industrial bodies, and government agencies.
A closed-loop algorithm should possess certain characteristics, such as:
\begin{itemize}
    \item \textit{Scalability}: The algorithm should have the capability to scale and accommodate the numerous devices and applications anticipated to connect to the network.
    \item \textit{Reliability}: The algorithm should operate dependably despite network failures and congestion.
    \item \textit{Security}: The algorithm can protect the network against security risks, including denial-of-service attacks and network slice isolation.
    \item \textit{Efficiency}: The algorithm should utilize resources efficiently.
\end{itemize}

The present study aims to propose an in-depth examination and enhancement of a scalable proactive closed-loop algorithm, \textbf{{\PSA}} - Proactive Close Loop Algorithm for Network slice Assurance, with a focus on proactive characteristics for service assurance in 5G and B5G networks enabled with network slices.
The algorithm is designed to optimize the utilization of network resources while ensuring compliance with the QoS requirements of multiple network slices operating in parallel.
Additionally, the {\PSA} is designed with flexible parameters that facilitate seamless adaptation to variable conditions and diverse network resources, thereby enhancing its performance across a range of scenarios and contributing to its multi-functionality in addressing disparate QoS requirements.

The remainder of this paper is organized as follows. Section \ref{sec:related_work} reviews related work on service assurance and resource allocation in network slices.
Section \ref{sec:e2e_orcherstration} provides a concise overview of the E2E network slice architecture.
Section \ref{sec:proactive_closed_loop_design} details the design and implementation of {\PSA}.
Section \ref{sec:experiment} presents a comprehensive performance evaluation of {\PSA} using a realistic simulation environment.
Finally, Section \ref{sec:conclusion} concludes the paper, discusses the benefits and future research directions.


\section{Related works}
\label{sec:related_work}

The automation of resource management and service assurance in next-generation networks has been extensively studied, particularly in the context of 5G and network slicing.
This entails the efficient management of network resources, and numerous algorithms have been proposed to dynamically allocate compute resources to VNF instances while optimizing network performance.
The primary objective is to meet SLAs while simultaneously minimizing resource utilization, operational costs, and energy consumption.
A central challenge lies in the dynamic and intelligent allocation of resources to ensure QoS across heterogeneous and customized network slices.

Early approaches to service assurance primarily relied on manual configuration and reactive mechanisms.
However, the increasing complexity of modern networks has necessitated the adoption of closed-loop automation, in which systems can autonomously monitor, analyze, plan, and execute (MAPE) actions. To efficiently manage compute resources in 5G networks, several algorithms and frameworks have been developed to extend the MAPE loop.
For example, Ren \textit{et al.} \cite{ren2016} proposed a distributed closed-loop architecture for real-time orchestration and service assurance.
This work emphasizes a hierarchical control plane and a knowledge-based service assurance system.
In addition, the DASA algorithm in their work addresses dynamic resource allocation for 5G network slices. Similarly, Ali \textit{et al.} \cite{ali2023proactive} introduced a service-assurance-based closed-loop framework for managing virtualized networks.
These methods, however, often require that a Key Performance Indicator (KPI) threshold be violated before any corrective action is taken, which is insufficient for applications with stringent latency and reliability requirements.
Other notable works, such as the Adaptive Service Assurer (ASA) \cite{ren_2018} and Govindarajan \textit{et al.} \cite{govindarajan2022closed}, also rely on reactive adjustments to maintain service levels.
While these frameworks optimize resource allocation, they predominantly operate in a reactive mode, responding only after a performance issue has occurred.
This reactive nature limits their capacity to anticipate fluctuations in network demand or proactively prevent service degradation.
Therefore, there is a clear need for predictive and intelligent mechanisms that can dynamically allocate resources in advance, ensuring continuous compliance with QoS requirements and supporting the strict performance demands of modern 5G and beyond networks.

Considering the literature on cellular networks, various categories of "slicing problems" have received attention and exploration.
One of the central areas that emerges is \textit{the allocation challenge resources} for physical nodes among network slices, including allocation of resource blocks within the CN and RAN \cite{boutaba2021ai, schardong2021nfv}.
The complexity of VNF resource allocation in SA is primarily characterized by the optimization of existing resources in a manner that satisfies the diverse requirements of distinct network slices \cite{wang2023artificial}. These requirements encompass, but are not limited to, the following:
\begin{itemize}
    \item \textit{Resource scarcity}: competition for limited resources, such as computing power, memory, storage, and network bandwidth, is crucial among network slices.
    Efficient resource allocation is of extreme importance to achieve optimal performance and avoid resource conflicts.
    \item \textit{QoS requirements}: network slices may exhibit diverse QoS requirements, which include factors such as latency, throughput, reliability, and availability.
    In order to fulfill the SLAs for each slice, it is essential that resource allocation takes into account these particular requirements.
    \item \textit{Dynamic resource demands}: the demands for resources may vary dynamically depending on factors such as network traffic patterns, user behaviours, and application requirements.
    Thus, the adaptability and real-time responsiveness of the resource allocation mechanism are important.
    \item \textit{Multi-dimensional resource optimization}: the process of resource allocation within the context of 5G network slice entails the simultaneous optimization of various dimensions, including but not restricted to CPU utilization, memory usage, power consumption, and network bandwidth.
    Therefore, the optimization problem for satisfying the QoS requirements for multiple slices while maintaining a balance between the different dimensions is a complex challenge.
    \item \textit{Isolation and security}: ensuring proper isolation between network slices is crucial to prevent interference, unauthorized access, and data breaches.
    Additionally, maintaining robust security measures within each network slice is imperative to protect sensitive information and mitigate potential vulnerabilities.
    \item \textit{Resource allocation policies}: to ensure optimal performance, it is essential to design resource allocation policies that effectively allocate resources based on the specific needs (such as those derived from SLA) and priorities of each network slice.
    This requires considering factors such as QoS requirements, traffic demands, latency constraints, and dynamic resource allocation.
\end{itemize}

In the context of network slicing, certain research endeavours have employed closed-loop mechanisms to address the associated challenges, such as \cite{nai2022, 10041293}.
The majority of existing research on network slice embedding has focused on addressing the one-shot optimization problem, which involves optimizing resource allocation based on average and/or static demands.
However, the latest evolution of the primary objective of SA is to dynamically and in real time allocate resources to across network slices or network installations in order to meet SA requirements while minimizing resource usage.
Furthermore, a critical limitation in much of the existing literature is a focus on isolated network segments.
Many proposals address resource management within the Radio Access Network (RAN) or core network (CN) or the transport network (TN) in isolation \cite{thaliath2022predictive, 10121746, vittal2023revamping}.
This fragmented approach fails to account for the holistic, end-to-end (E2E) performance of a network slice, where performance bottlenecks can arise at any point along the service chain.
An effective solution must be capable of orchestrating resources across the entire E2E path to guarantee a seamless and consistent user experience.

In contrast to these existing efforts, our proposed {\PSA} introduces a novel approach that overcomes these limitations.
While prior closed-loop systems are predominantly reactive, {\PSA} is inherently proactive.
It leverages a forecasting model to anticipate future resource demands and potential network congestion, enabling the system to scale resources before performance degradation occurs.
This predictive capability allows {\PSA} to maintain high levels of QoS, even under rapidly changing network loads.
Moreover, unlike solutions focused on individual network segments, {\PSA} provides end-to-end network orchestration. 
Our work advances the field of network slice assurance by developing a proactive closed-loop algorithm that integrates machine learning for dynamic resource allocation across end-to-end 5G network slices.
In contrast to Marinova \textit{et al.}~\cite{marinova2025e2e}, whose research presents a holistic framework for E2E network slice assurance through data collection, MLOps, and multi-domain closed-loop control with a focus on system architecture and operational workflows, our approach emphasizes algorithmic innovation for predictive scaling and SLA adherence in dynamic network environments.
This focus enables real-time adaptation to traffic variations within network slices, reducing KPI violations and optimizing resource utilization.
It intelligently allocates and scales resources across the entire E2E network slice, including both the core and transport domains.
By adopting this holistic perspective, {\PSA} ensures service assurance across the entire service chain, making it more robust and effective in managing the complexities of modern, virtualized 5G and B5G networks.
The comparative analysis presented in our evaluation section will further highlight the significant performance improvements achieved by our proactive and end-to-end approach.

\section{End-to-End Orchestration}
\label{sec:e2e_orcherstration}

The trend of network softwarization involves an extensive redesign of the creation, implementation, deployment, management, and maintenance of network equipment and components through the use of software programming.
This approach leverages the inherent characteristics of software, such as flexibility and rapid design, development, and deployment, throughout the whole life cycle of network equipment and components.
Two distinct architectures for the 5G core network have been established by the 3rd Generation Partnership Project (3GPP), namely the reference point architecture and the service-based architecture \cite{5g-service-requirements}.
Within the context of the reference point architecture, a distinct reference point is established between two distinct network functions, thereby enabling the functions to interact in communication with one another via these reference points. 

Throughout a service-based architecture, identical interfaces are allocated to corresponding functionalities across all interfaces.
One of the defined aspects of 5G-CN by 3GPP is decoupling the user plane function (UPF) and control plane function.
Through this approach, the novel architecture is able to achieve flexibility, efficacy, and scalability in both the development and operation of 5G/B5G networks.
Conversely, the system has the ability to enhance resource allocation through the utilization of traffic patterns and demands. 
The control plane function provides the ability to dynamically deliver and distribute resources, including radio bearers and QoS parameters. 
In contrast, the UPF prioritizes the optimization of data transmission efficiency.

\begin{figure}[ht]
    \centering
    \includegraphics[width=0.99\columnwidth]{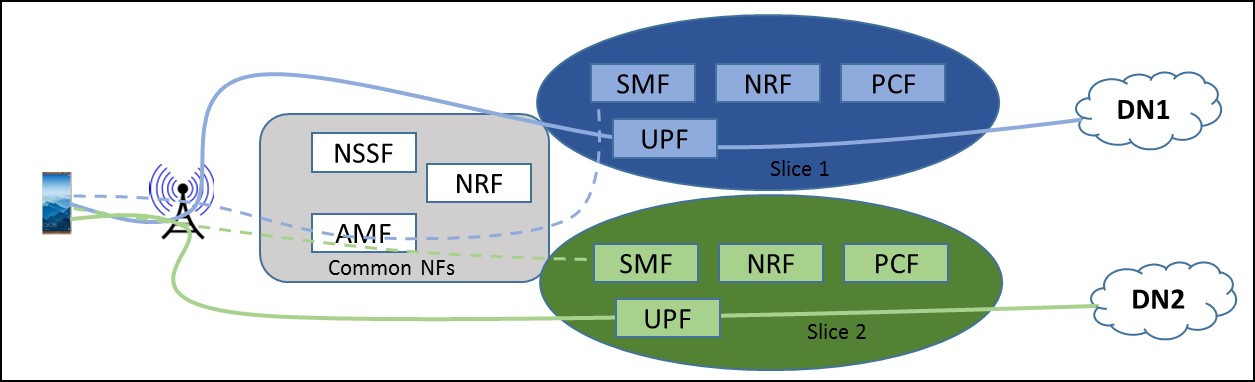}
    \caption{A reference E2E network slice concept  \cite{Network-e2e-3gpp}}
    \label{fig:reference-e2e-network-slicing-concept-3gpp}
\end{figure}

With the new design mentioned above, the concept of E2E orchestration has recently gained prominence as an innovative concept in the domain of 5G and B5G networks \cite{afolabi2018network}.
Orchestration refers to the comprehensive management as well as coordination of multiple network functions, resources and services across the network infrastructure, resulting in parallel degrees of flexibility, efficiency and automated operation. 
Demonstrated in Fig. \ref{fig:reference-e2e-network-slicing-concept-3gpp} from 3GPP, the implementation of end-to-end orchestration provides a holistic strategy for handling network operations, supporting operators to efficiently manage and enhance all network components, ranging from the RAN, the TN, to the CN.
The E2E network slice infrastructure employs various management domains and utilizes modern SDN and NFV technologies to facilitate flexible resource allocation, service chaining, and policy enforcement.
Thus, an E2E network slice involves a physical infrastructure comprising network, computing, and storage resources that are programmable and embedded throughout the end-to-end communication paths.
The study \cite{8802040} provides a deep overview of E2E network slice in both vertical and horizontal directions with a detailed discussion on network slice isolation, and application use cases that enable a comprehensive infrastructure for network slice in a 5G network.
Based on the showcases, we can indicate the significance of network slice isolation, which guarantees the independent and secure operation of each network slice, without any external factors or interference from other network slices.
Ensuring the confidentiality, integrity, and efficiency of each network slice is of the highest priority in situations where sensitive or vital applications are utilized, thereby emphasizing the significance of network slice isolation. 
Thus, the mechanisms and techniques need to be revisited to create network slice isolation effectively and to address various challenges that arise in this scenario, including resource allocation, traffic management, and security enforcement.
As per the definition provided in reference \cite{alliance2016description}, the concept of network slice comprises three distinct layers.
\begin{itemize}
    \item The \textit{Service Instance Layer} refers to the provision of services to end-users or businesses that are supported. A service instance is the representation of each individual service.
    \item The \textit{Network Slice Instance Layer} covers the various network slice instances that are available for provisioning. A network slice instance is responsible for delivering the necessary network functionalities to support the service instance.
    \item The \textit{Resource Layer} is responsible for providing all requisite virtual or physical resources and network functions essential for the instantiation of a network slice.
\end{itemize}

Despite the numerous benefits that E2E network slice offers for 5G and B5G networks, there remain certain gaps in knowledge and research opportunities \cite{8685766} such as RAN virtualization and network slice, holistic and intelligent network slice orchestration, secure network slices, and quality of services in multiple network slices.
Drawing from the previously mentioned review, below we will construct a 5G E2E network slice architecture in a simulator environment, with the goal of implementing and addressing intelligent network management in the context of service assurance.
\begin{figure}[ht]
    \centering
    \includegraphics[width=1\columnwidth]{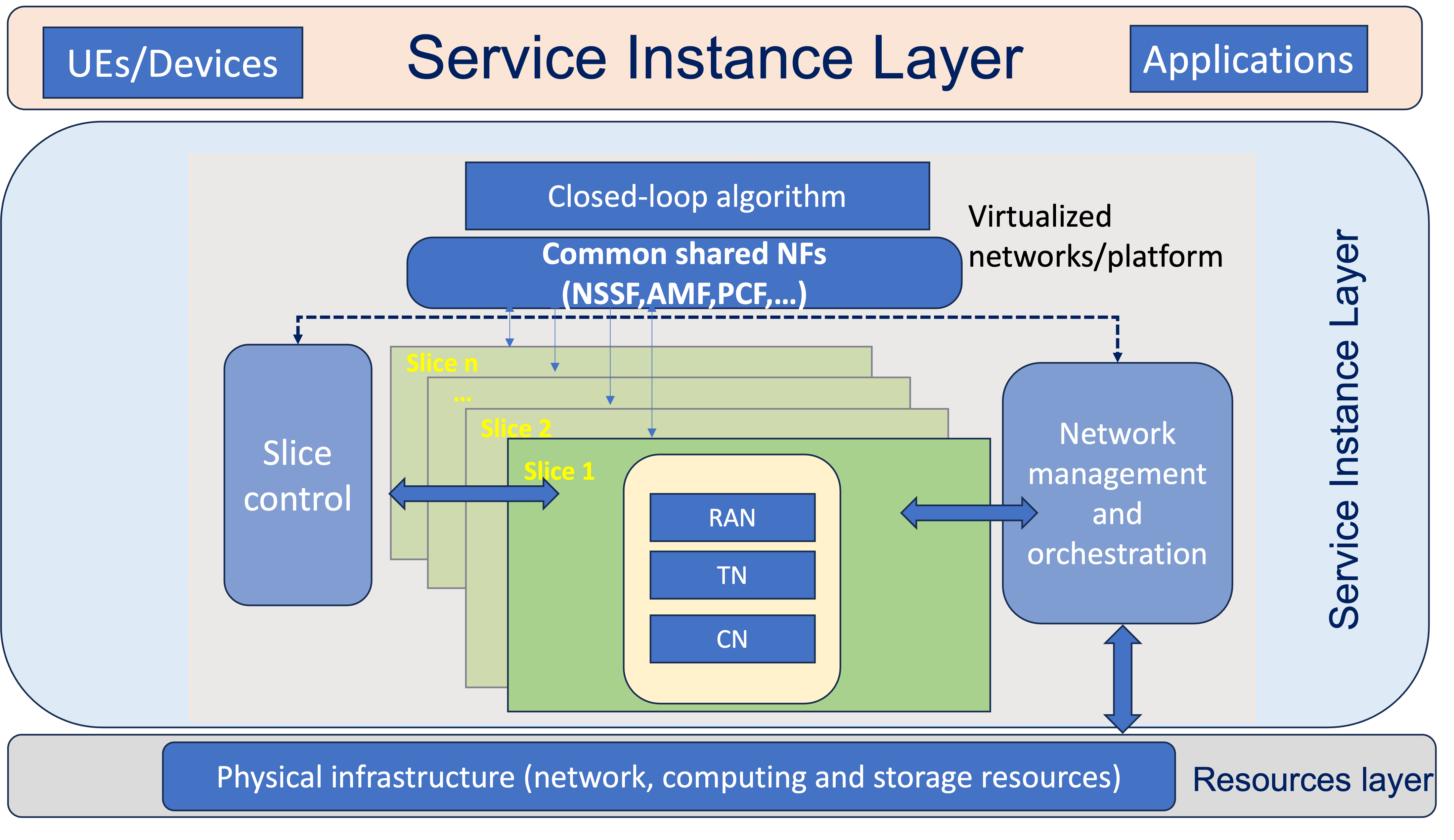}
    \caption{End-to-end network slice orchestration with closed-loop algorithm}
    \label{fig:e2e-slice-orchestration}
\end{figure}
Fig. \ref{fig:e2e-slice-orchestration} depicts a high-level view of our E2E network slice infrastructure with a closed-loop algorithm on a 5G network.

It consists of five primary components, including network slice control, MANagement and Orchestration (MANO), virtualized networks/platforms, physical infrastructure, and a closed-loop algorithm.
In the virtualized networks/platform, there exists a set of commonly shared network functions (NFs) \cite{Hua5gns, etsi23next}, including but not limited to the Network Slice Selection Function (NSSF), Policy Control Function (PCF), and Access and Mobility Management Function (AMF).
This approach offers several advantages, such as cost savings on hardware and software, enhanced network efficiency via a reduction in the number of VNF instances that must be deployed, and increased network scalability by facilitating the creation of new network slices.
In addition, network slice control is used to establish and manage network slices, enforce network slice policies, and monitor slice performance.
In cooperation with network slice control, the MANO component is in charge of ensuring optimal network performance and functionality. This includes facilitating network visibility, equipping network administrators with effective management tools, and automating network management processes \cite{8406279}.
Next, the integration of a closed-loop algorithm has been implemented with the goal of improving coordination between the network slice control and MANO components. 
This integration has enabled the management of network slice in reliable and efficient ways, while also aligning with customer requirements and satisfying QoS.
Finally, the aforementioned components are in charge of controlling and managing a shared physical infrastructure in order to establish an E2E network that optimizes the entirety of the network capability, from the RAN, the TN and the CN.
Thus, every E2E network slice is created with an isolated virtual network, a set of VNF instances, dedicated virtual computing and storage resources, with several shared common NFs.


\section{Proactive closed-loop algorithm design}
\label{sec:proactive_closed_loop_design}

\subsection{Resource model}

Each network slice $\SliceIdx$ ( $\SliceIdx \in \SliceSet$, where $\SliceSet$ is the set of network slice instances) can utilize multiple VNF instances $\VNFIdx$ ($\VNFIdx \in \VNFSet_{\SliceIdx}$, where $\VNFSet_{\SliceIdx}$ is the set of VNF instances for slice $\SliceIdx$).
Each VNF instance $\VNFIdx$ requires resources $\ResourceIdx$ ($\ResourceIdx \in \ResourceSet = \{\CPU, \RAM, \STO\}$ with capacity $\CAPA_{\ResourceIdx}^{\VNFIdx}$ and untilization $\ResourceUtilizationVector_{\ResourceIdx}^{\VNFIdx}$.
In the network, there is a set of physical machines ($\SetPhyMachine$) that have resource capacities $\CAPA_{\ResourceIdx}^{\PMIdx}$, where $\PMIdx \in \SetPhyMachine$.
Note that VNF instances are instantiated in physical machines
through the virtualization platform, and each of them has an
amount of $\CAPA^{\PMIdx}$. 
At all times, the resources utilized across all network slices should not exceed those provided by the physical
machines:
\begin{equation}
\label{eq:check_total_slice_cap_with_phy}
\sum_{\SliceIdx \in \SliceSet} \sum_{\VNFIdx \in \VNFSet_{\SliceIdx}} \CAPA^\VNFIdx_\ResourceIdx  \leq \sum_{\PMIdx \in \PMSet} \CAPA^\PMIdx_\ResourceIdx,  \forall \ResourceIdx \in \ResourceSet
\end{equation}

For instance, consider a CN with a data center configuration of 2 \SetPhyMachine. 
Each \PhyMachineIdx \space has a capacity of 2 \CPU(s), 3GB of \RAM, and 5GB of \STO. 
At any time, the total amount of resources used on network slices, allocated to VNF instances, must not exceed 4 \CPU(s), 6GB of \RAM, and 10GB of \STO, according to Formula \eqref{eq:check_total_slice_cap_with_phy}.

At any timestamp, the total link capacity across network slices should not exceed the link capacity provided by the network and must satisfy the equation \eqref{eq:constrant_link_capacity}.
Assuming that we have access only to the links that are connected to the core network. At any given time, each network slice $\SliceIdx$ requires a link instance.
Each link instance is allocated a specific amount of resources denoted by $\LinkIdx_{\SliceIdx}$, and there is also a collection of physical links $\LinkIdx^{\PHY} \in \LinkSet^{\PHY}$.
Consequently, the total link capacity allocated for network slices mustn't exceed the total physical link capacity provided by the network infrastructure at any given time, as denoted by Formula \eqref{eq:constrant_link_capacity}:

\begin{equation}
    \label{eq:constrant_link_capacity}
    \sum_{\SliceIdx \in \SliceSet} \LinkCapacity \leq \CAPA_{\LinkIdx^{\PHY}}
\end{equation}
where $\LinkCapacity$ is the virtual link capacity and $\CAPA_{\LinkIdx^{\PHY}}$ is the physical link capacity provided by network infrastructure.
Formula \eqref{eq:constrant_link_capacity} is utilized to verify the link configuration at any point within the network, from the RAN to the CN and from the CN to the data network, where the algorithm is executed.

\subsection{Implementation of {\PSA}}

\begin{figure}[ht]
    \centering
    \includegraphics[width=1\columnwidth]{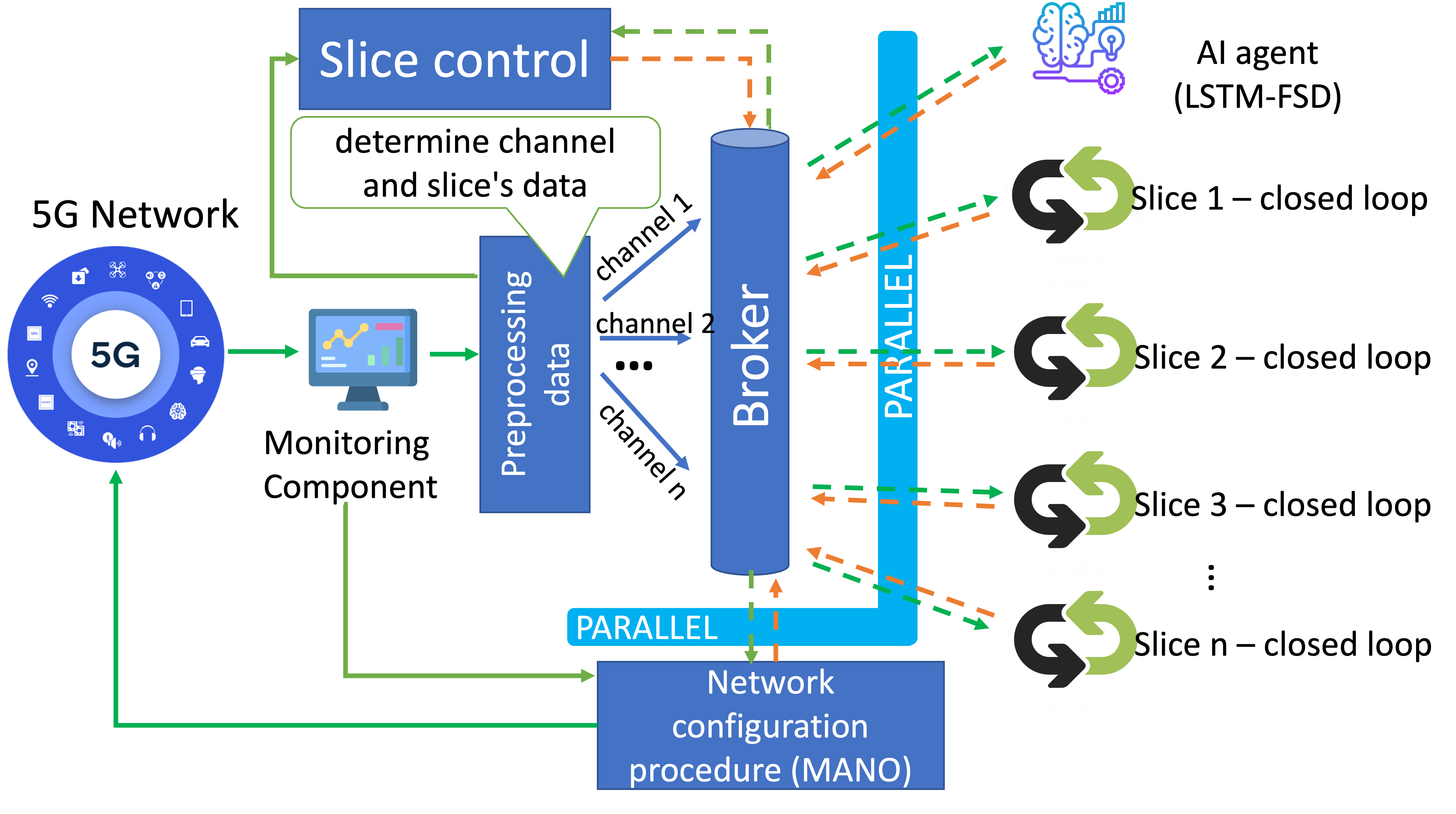}
    \caption{Proactive closed-loop architecture}
    \label{fig:proposal-proactive-closed-loop-architect}
\end{figure}

The proposed architecture in Fig. \ref{fig:proposal-proactive-closed-loop-architect} leverages the closed-loop algorithm at the network slice level to enable parallel SA processing and minimize the complexity of the algorithm in development and scalability. This approach enhances the network design's ability to expedite the processing and execution of actions.
At each time window {\TimeWindow}, {\PSA} estimates the network slice resource consumption per throughput unit (e.g., Mbps, Gbps) as follows:
\begin{equation}
\label{eq:estimate_resources_per_th_unit}
    \RequireComputeResourcePerThroughput = \sum\limits_{\VNFIdx \in \VNFSet_{\SliceIdx}} \sum\limits_{\ResourceIdx \in \ResourceSet^{\VNFIdx}} \frac{ \CAPA_{\ResourceIdx} \cdot \ResourceUtilizationVector_{\ResourceIdx}}{\Throughput_{\SliceIdx}}
\end{equation}
where $\Throughput_{\SliceIdx}$ is the throughput of a given network slice.
By utilizing the network slice resource consumption per throughput, {\PSA} is capable of efficiently calculating and identifying changes in traffic load. 
Therefore, it can effectively adjust resource allocation in response to fluctuations in traffic load, whether they involve an increase or a decrease in resources.
We can select the quantity of current VNF instances assigned to a specific network slice by taking the total resources configuration of a given network slice and dividing by the maximum physical resources per VNF instance.
The required number of VNF instances needed for a given network slice during the upcoming time window can be computed as follows:
\begin{equation}
    \label{eq:determine_number_of_vnf_instances}
    \NumberOfVnf_{\SliceIdx} = \ceil*{\max\limits_{\ResourceSet} 
              \left\{ \frac{\sum\limits_{\VNFIdx \in \VNFSet_{\SliceIdx}} 
                            \sum\limits_{\VNFInstance}
                           \sum\limits_{\ResourceIdx \in \ResourceSet^{\VNFIdx}}
                           \CAPA_{\ResourceIdx}^{\VNFInstance}}
                           {\MaxCAPVNFSlice}
              \right\}}
\end{equation}
where {$\MaxCAPVNFSlice$} is the maximum allowed resource capacity per VNF instance that could be instantiated in a network slice.

Relying upon the results derived from Formula \eqref{eq:estimate_resources_per_th_unit}, {\PSA} is able to compute the amount of resources required to facilitate the processing of a single throughput unit. Subsequently, this information can be utilized in combination with a machine learning (ML) agent model to forecast the amount of compute resources necessary for a given network slice.
Thus, with the assistance of an ML agent, we can predict the throughput at the next time step. 
This time step can be configured as $\TimeWindow+1$, or $\TimeWindow+n$, and {\PSA} can estimate the amount of compute resources and link resources $\SliceConfig$ needed for a given network slice. 
Please note that $\SliceConfig$ is a vector formed by combining compute and link resources (creating a higher dimension vector) defined as: 
\begin{equation}
    \label{eq:calculate_slice_resources}
    \SliceConfig = (\PredictedThroughput_{\SliceIdx} \cdot \RequireComputeResourcePerThroughput, \TrafficBoundary)  
\end{equation}
where:
\begin{itemize}
    \item $\PredictedThroughput_{\SliceIdx}$: Predicted throughput in the time window $\TimeWindow + 1 $ (or $\TimeWindow + n $ depending on the model configuration).
    \item $\TrafficBoundary$: throughput boundary obtained by a traffic prediction model, see next paragraph for clarification and \eqref{eq:traffic_boundary} for its value.
\end{itemize}
 
To determine the predicted throughput $\PredictedThroughput_{\SliceIdx}$ of a given network slice, a machine learning framework, $\lpkpi$~\cite{del_2023}, was used to analyze the historical data.
The $\lpkpi$ framework comprises two components: the LSTM-FSD model, which performs short-term throughput forecasting using traffic, resource utilization, and network slice configuration data; and the LP-KPI model, which employs an ILP-based approach to predict additional KPIs, such as delay and packet loss, by integrating the predicted throughput with the current network state.
Subsequently, the model was deployed in conjunction with {\PSA} to get the predicted traffic and estimate network KPIs in the upcoming time.
It is important to acknowledge that the ML agent is not capable of guaranteeing 100\% accuracy. 
Therefore, {\PSA} will implement a monitoring interval $\MonitorTrafficTime$ to prevent abnormal traffic or incorrect prediction, and avoid excessive scaling.
During the monitoring interval $\MonitorTrafficTime$, an error rate $\TrafficErrorRate$ will be computed at each timestamp that is utilized to establish the traffic boundary.
This error rate can be calculated by determining the mean of the absolute differences between the predicted and actual throughput for each time step within the monitoring window, $\MonitorTrafficTime$.
Through the implementation of this methodology, it is possible to maintain a more consistent prediction of traffic fluctuations and scaling procedures.
Consequently, a traffic boundary based on a forecasting model was defined as follows:
\begin{equation}
    \label{eq:traffic_boundary}
    \TrafficBoundary = \TrafficModelAccuary \cdot \PredictedThroughput_{\SliceIdx} + \TrafficErrorRate
\end{equation}

\begin{itemize}
    \item $\TrafficModelAccuary$: is the accuracy of the traffic prediction model. For example, the accuracy of the \lstmfsd \space model.
    \item $\TrafficErrorRate$: is the average error rate between the actual and the predicted traffic in the monitoring time window $\MonitorTrafficTime$.
\end{itemize}

In addition, the algorithm possesses the capacity to calculate and dynamically allocate link resources for individual network slices within the network, contingent on the traffic load forecast (refer to lines 23 to 32 in Algorithm \ref{alg:proactive_closed_loop} for further details).
The specifics of our {\PSA} are delineated in Algorithm \ref{alg:proactive_closed_loop}.
This algorithm is designed to run parallel instances across different network slices, thereby enabling efficient resource allocation and scaling for network slices.
Utilizing a closed-loop with the ML approach, the system can proactively allocate resources in response to changing network conditions, thereby optimizing performance and reducing resource under-utilization.
Thus, this leads to considerable enhancements in network efficiency and dependability.
\begin{figure}[!t]
    \centering
    \includegraphics[width=2.3\columnwidth, angle=90]{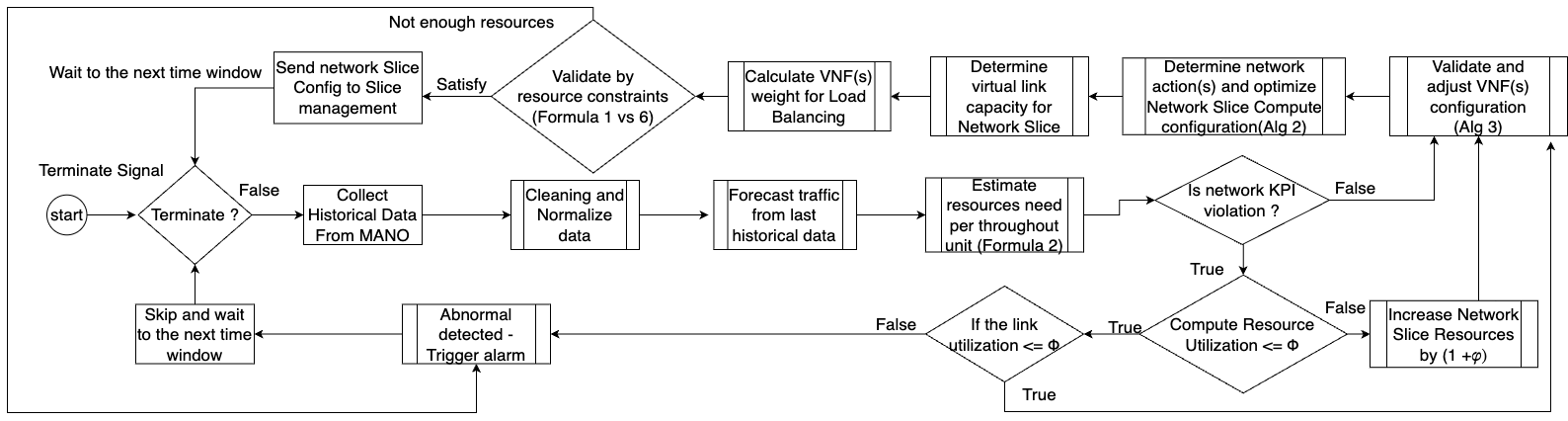}
    \caption{Overview of {\PSA}: a high-level flowchart.}
    \label{fig:high_level_psa}
\end{figure}

\begin{algorithm}[!t]
    \caption{PROACTIVE CLOSED-LOOP ALGORITHM}
    \label{alg:proactive_closed_loop}
    \hspace*{\algorithmicindent} \textbf{Input:} Network slice configuration parameters (Table \ref{tab:proactive_closed_parameters}) \\
    \hspace*{\algorithmicindent} \textbf{Output:} Network slice configuration, network action set
    \begin{algorithmic}[1]
        \State $signal^{terminate} \gets False$
        \For{$\forall \TimeWindow$ AND $signal^{terminate}$ is $False$}
            \State $\HistoricalData \gets$ MANO by $\NumberOfSampling$ sampling data
            \State $\HistoricalDataNormalized \gets pre\_process(\HistoricalData)$
            \Comment{Clean and normalized historical traffic and network slice configurations.}
            \State $\PredictedThroughput \gets \lstmfsd$ agent by $\HistoricalDataNormalized$
            \State $\TrafficBoundary \gets$ Formula \eqref{eq:traffic_boundary}
            \For{every timestamp $t_{i} \in t$}
                \State $temp \gets$ Formula \eqref{eq:estimate_resources_per_th_unit}
                \State $\RequireComputeResourcePerThroughput \gets \max \{ \RequireComputeResourcePerThroughput, temp \}$
            \EndFor
            \State $\SliceConfig \gets$ Formula \eqref{eq:calculate_slice_resources}, \ScalingActionSet $\gets \emptyset$
            \State $is\_kpi\_violation \gets $ check\_network\_KPIs($\TargetSliceKPIs$)
            \If{$is\_kpi\_violation$ AND $\forall \ResourceUtilizationIdx \leq \ratioResUsage$; $\ResourceIdx \in \ResourceSet^{\VNFIdx}$ }
                \If{$\LinkUtilization \leq \ratioResUsage$}
                \Comment{$\LinkUtilization$: Link capacity utilization}
                    \State Send $signal^{alarm}_{abnormal}$ to MANO
                    \State \textbf{Skip and wait for next time window}
                \EndIf
            \ElsIf{$\forall \ResourceUtilizationIdx > \ratioResUsage$, $\ResourceIdx \in \ResourceSet^{\VNFIdx}$}
                \State $\SliceConfig \gets (1 + \ratioOverProvision) \cdot \SliceConfig$
            \EndIf
            \State $\SliceConfig \gets$ Algorithm \eqref{alg:validate_algorithm}
            \State $\ScalingActionIdx^{\COMP}, \VNFSet_{\SliceIdx} \gets$ Algorithm \eqref{alg:scaling_algorithm}
            \Comment{$\ScalingActionIdx$: network action, \COMP: compute resources.}
            
            \State $\ScalingActionIdx^{\LinkIdx} \gets No\_action$ 
            \If{$\LinkCapacity > \TrafficBoundary$ AND $\LinkUtilization < \ratioResUsage$} 
                \State $\ScalingActionIdx^{\LinkIdx} \gets scale\_down\_link$
                \State $s^{\LinkIdx} \gets \TrafficBoundary$ 
            \ElsIf{$\LinkUtilization > \ratioResUsage$}
                \State $\ScalingActionIdx^{\LinkIdx} \gets scale\_up\_link$
                \State $s^{\LinkIdx} \gets \max \{ s^{\LinkIdx} \cdot \ratioScaling, \TrafficBoundary \}$
            \Else
                \State $s^{\LinkIdx} \gets \LinkCapacity$
            \EndIf

            \State $\ScalingActionSet \gets \ScalingActionIdx^{\COMP} \cup \ScalingActionIdx^{\LinkIdx}$
            \State $enough\_resource \gets $ validate by Formula \eqref{eq:check_total_slice_cap_with_phy} AND \eqref{eq:constrant_link_capacity}
            \If{not $enough\_resource$}
                \State Send $signal^{alarm}_{resources}$ to MANO
            \Else
                \State Calculate VNF(s) weights for the load balancer
                \State Send $signal^{apply}_{resources}$ update configurations ($\VNFSet_{\SliceIdx} \cup s^{\LinkIdx}, \ScalingActionSet$) for network slice to Network Slice manager
            \EndIf
        \EndFor
        
        \State \textbf{Return}
    \end{algorithmic}
\end{algorithm}
\begin{algorithm}[h]
    \caption{SCALING ALGORITHM}\label{alg:scaling_algorithm}
   
    \hspace*{\algorithmicindent} \textbf{Input:} new compute resources configuration $\SliceResourceConfig$ \\
    \hspace*{\algorithmicindent} \textbf{Output:} $\ScalingActionIdx^{\COMP}$, set of VNF(s) configuration  $\VNFSet_{\SliceIdx}$
    \begin{algorithmic}[1] 
         \State $V_{\SliceIdx}^{\VNFIdx} \gets$ stack of current VNF(s) configurations for network slice \SliceIdx
         \State $\CAPA_{\SliceIdx}^{current} \gets \sum \limits_{\VNFIdx \in V_{\SliceIdx}^{\VNFIdx}} \sum \limits_{\ResourceIdx \in \ResourceSet^{\VNFIdx}} \sum \limits_{\VNFInstance} \CAPA^{\VNFInstance}_\ResourceIdx$
         \State $\ratioOverProvision_{current} \gets \max\{\frac{\CAPA_{\SliceIdx}^{current} - \SliceResourceConfig}{\CAPA_{\SliceIdx}^{current}}\}$
         \Comment{Take the maximum over-provisioning ratio of compute resources}
         
         \If {$\forall \ResourceIdx \leq \ResourceIdx_{i}$, $r \in \SliceResourceConfig, \ResourceIdx_{i} \in \CAPA_{\SliceIdx}^{current}$ AND $\ratioOverProvision_{current} \leq \ratioOverProvision$}
            \State \textbf{Return} $\ScalingActionIdx^{\COMP} = no\_action$, $V_{\SliceIdx}^{\VNFIdx}$
         \EndIf
         
         \State $\NumberOfVnf^{deployed} \gets$ Count $\VNFInstance \in V_{\SliceIdx}^{\VNFIdx}$
         
         \State $\NumberOfVnf \gets$ Formula \eqref{eq:determine_number_of_vnf_instances} using $\SliceResourceConfig$

        \State $v \gets V_{\SliceIdx}^{\VNFIdx}.pop()$ 
        
         \If{$\NumberOfVnf = \NumberOfVnf^{deployed}$}
            \State $R^{\COMP} \gets \abs{\SliceResourceConfig - \CAPA_{\SliceIdx}^{current}}$
            \Comment{$R^{\COMP}$: require resources.}
            \If{$\exists \CAPA_{\ResourceIdx} > \CAPA_{\ResourceIdx_{i}}, \forall r \in R^{\COMP}, \ResourceIdx_{i} \in \ResourceSet^{v}$} \Comment{ $\ResourceSet^{v}$: resources of VNF $v$.}
                \State $\ScalingActionIdx^{\COMP} \gets scale\_up$
                \State $m \gets \ratioScaling \cdot \ResourceSet^{v}$
                \State $\ResourceSet^{v} \gets \max\{\MinCAPVNFSlice, m, \ResourceSet^{\COMP}\}$
            \Else
                \State $\ScalingActionIdx^{\COMP} \gets scale\_down$
                \State $\ResourceSet^{v} \gets \max\{\MinCAPVNFSlice, R^{\COMP}\}$
            \EndIf
            \State $V_{\SliceIdx}^{\VNFIdx}.push(v)$
            
        \ElsIf{$\NumberOfVnf > \NumberOfVnf^{deployed}$}
            \State $p \gets \min\{\frac{\MaxCAPVNFSlice - \ResourceSet^{v}}{\ResourceSet^{v}}\}$
            \State $\ScalingActionIdx^{\COMP} \gets scale\_out$
            \State $scale\_up$  $\ResourceSet^{v}$ by $p$ percent
            \State $V_{\SliceIdx}^{\VNFIdx}.push(v)$
            \State $\ResourceSet^{new} \gets \max\{\MinCAPVNFSlice, \ResourceSet^{\COMP} - \sum\limits_{\VNFIdx \in V_{\SliceIdx}^{\VNFIdx}} \sum \limits_{\ResourceIdx \in \ResourceSet^{\VNFIdx}} \sum \limits_{\VNFInstance} \CAPA_{\ResourceIdx}^{\VNFInstance}\}$ \Comment{$\ResourceSet^{new}$: resources for new VNF instance \xspace $v^{new}$.}
            \State $V_{\SliceIdx}^{\VNFIdx}.push(v^{new})$
        \Else
            \State $p \gets min\{\frac{\sum\limits_{\VNFIdx \in V_{\SliceIdx}^{\VNFIdx}} \sum \limits_{\ResourceIdx \in \ResourceSet^{\VNFIdx}} \sum \limits_{\VNFInstance} \CAPA_{\ResourceIdx}^{\VNFInstance} - \SliceResourceConfig}
            {\sum \limits_{\ResourceIdx \in \ResourceSet^{V_{\SliceIdx}^{\VNFIdx}[Last]}} \CAPA^{\VNFInstance}_{\ResourceIdx}}\}$
            \State $\ScalingActionIdx^{\COMP} \gets scale\_in$
            \State $v \gets V_{\SliceIdx}^{\VNFIdx}.pop()$
            \State Decrease $\ResourceSet^{v}$ by $p$ percent
            \State $\ResourceSet^{v} \gets \max\{\ResourceSet^{v}, \MinCAPVNFSlice\}$
            \State $ V_{\SliceIdx}^{\VNFIdx}.push(v)$
        \EndIf
        \State \textbf{Return} $\alpha^{\COMP}, V_{\SliceIdx}^{\VNFIdx}$
    \end{algorithmic}
\end{algorithm}
\begin{algorithm}[h]
    \caption{VALIDATE NETWORK SLICE CONFIGURATIONS}\label{alg:validate_algorithm}
   
    \hspace*{\algorithmicindent} \textbf{Input:} Network slice configurations $\SliceConfig$, slice's historical data and KPI thresholds \\
    \hspace*{\algorithmicindent} \textbf{Output:} Validated network slice configurations
    \begin{algorithmic}[1]
        \State $has\_kpi\_violation \gets True$
        \While{$has\_kpi\_violation$}
            \State $has\_kpi\_violation \gets$ validate $\SliceConfig $ by $\lpkpi$ framework
            \If{$\exists kpi \in $ Slice's KPI thresholds is not satisfy}
                \State $\SliceConfig \gets \SliceConfig \cdot \ratioValidateScaling$
            \Else
                \State $has\_kpi\_violation \gets False$
            \EndIf
        \EndWhile
        \State \textbf{RETURN} $\SliceConfig$
    \end{algorithmic}
\end{algorithm}

\begin{table}[ht]
    \centering
    \resizebox{\columnwidth}{!}{
    \begin{tabular}{|l|c|c|}
        \hline
        Notation    & Definition & \makecell{Share \\among network slices} \\
        \hline
        $\ratioOverProvision$   & Accepted over-provisioning resource ratio &   \\ 
        $\ratioScaling$         & Minimum scaling step ratio                &   \\ 
        $\TrafficModelAccuary$  & Traffic prediction accuracy ratio         &  \cmark \\ 
        $\ratioResUsage$        & Expected resource utilization ratio       &   \\ 
        $\ratioValidateScaling$ & Resources validation scaling ratio        &  \\ 
        $\MaxCAPVNFSlice$       & Maximum allocated resources per VNF instance for network slice       &   \\
        $\MinCAPVNFSlice$       & Minimum allocated resources per VNF instance for network slice       &   \\
        $\NumberOfSampling$     & Number of sampling data                   &  \cmark \\
        $\totalPhyCap$          & Physical nodes configuration              &  \cmark \\
        $\totalLinkCap$         & Total physical link capacity              &  \cmark \\
        $\TimeWindow$           & Time window                               &  \cmark \\
        $\MonitorTrafficTime$   & Monitoring traffic time window            &   \\
        $\TargetSliceKPIs$      & Set of target network slice KPIs                  &  \\
        \hline
    \end{tabular}
    }
    \caption{{\PSA} parameters}
    \label{tab:proactive_closed_parameters}
\end{table}

\begin{table}[ht]
    \centering
    \resizebox{\columnwidth}{!}{%
    \begin{tabular}{|p{1cm}|c|l|p{3.5cm}|}
        \hline
        Domain & No. & Action ($\alpha$) & Description\\
        \hline
        \multirow{4}{4em}{Core Network} & 1 & $scale\_up$ & Scale-up network slice resources. \\
        & 2 & $scale\_down$ & Scale-down network slice resource. \\
        & 3 & $scale\_out$ & Add VNF instance(s). \\
        & 4 & $scale\_in$ & Remove VNF instance(s). \\
        \hline
        \multirow{2}{4em}{Transport Network} & 5 & $scale\_up\_link$ & Increase virtual link capacity. \\
        & 6 & $scale\_down\_link$ & Decrease virtual link capacity. \\
        \hline
        Both & 7 & $no\_action$ & No action needed. \\
        \hline
    \end{tabular}
    }
    \caption{{\PSA} actions}
    \label{tab:action_space}
\end{table}

As illustrated in the high-level flowchart in Figure \ref{fig:high_level_psa}, the {\PSA} functions by leveraging the aforementioned formulas embedded within the ML model.
Our proactive closed-loop algorithm in the network aims to provide an efficient and dynamic service deployment and management capable of maintaining the QoS of multiple network slices by detecting KPI violations quickly and accurately, taking corrective actions, and assigning appropriate resources to resolve KPI violations on time.
Thus, the algorithm has been developed with flexible parameters, shown in Table \ref{tab:proactive_closed_parameters}, enabling operators to customize {\PSA} by themselves in order to meet the specific requirements of the network slice and align it to their infrastructure.
To keep things simple, we split {\PSA} into two main algorithms:

In the first Algorithm \ref{alg:proactive_closed_loop}, the algorithm aims to forecast the upcoming traffic (per network slice), see lines 3 to 6.
By combining historical information with Formulas \eqref{eq:estimate_resources_per_th_unit} and \eqref{eq:calculate_slice_resources}, the algorithm estimates the resources required for the forthcoming timestamp, as shown in lines 7 to 11.
Before optimizing resources and performing scaling, the algorithm conducts a KPI violation check in line 12.
If all resources are below the $\ratioResUsage$ rate but have KPI violations, there is a possibility of an abnormal event in the system, such as a dropped link connection or loss of power in a node.
In such cases, the algorithm will trigger an alarm in the system.
In the event of a KPI violation, the algorithm will attempt to increase resources by a factor of $\ratioOverProvision$ to mitigate potential bottlenecks caused by insufficient resources.
Subsequently, the configuration of the network slice is evaluated using the $\lpkpi$ framework to estimate the necessary resources, with the objective of preventing KPI violations associated with the network slice.
Afterwards, the final network slice configuration will be processed by Algorithm \ref{alg:validate_algorithm}, which will perform precise network checks. The algorithm verifies the network infrastructure constraints and ensures that there is sufficient network capacity for the network slice before executing any operations.

The second Algorithm \ref{alg:scaling_algorithm} is used to optimize the network slice configuration obtained in the first phase and perform accurate actions, shown in Table \ref{tab:action_space}.
To mitigate the issue of frequent scaling, the algorithm utilizes a minimum scaling increment denoted by $\ratioScaling$. It is used to determine the minimum quantity of resources required in the event of infrastructure expansion.
During the process of scaling up or scaling down, the algorithm allocates resources for each compute resource type independently, in order to ensure that each resource type is in accordance with the upcoming traffic.
In the scaling-out phase, the algorithm tries to maximize the resources of the last VNF (in the same network slice) while maintaining a consistent ratio of values among compute resources.
Subsequently, the algorithm computes and adds a new VNF into the network slice only if the last VNF has exhausted its maximum allowable resources.
The aforementioned mechanism is also applicable to the scaling in phase but in the reverse direction. 
Consequently, the algorithm is capable of calculating the optimal resources necessary for VNFs in the subsequent period, aligning them with the network slice KPIs.


\section{Evaluation}
\label{sec:experiment}

This section will provide an overview of our experimental setup and showcase our service assurance algorithm designed to support 5G networks with network slices, encompassing diverse network slice categories.

\subsection{5G network slice environment}

\begin{figure}
    \centering
\includegraphics[width=1\columnwidth]{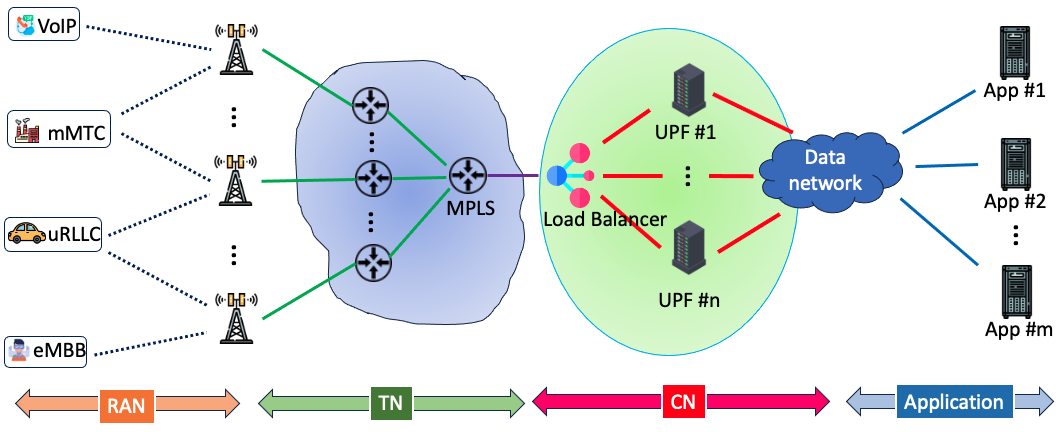}
    \caption{5G End-to-End network simulation topology}
    \label{fig:e2e_network}
\end{figure}

A packet-level simulation was developed using Omnet++ \cite{del_2022} to emulate a 5G network environment, incorporating support for slicing features.
The 5G E2E network slice simulation involves the initial configuration of four distinct network slices: uRLLC (video gaming), mMTC (IoT), eMBB (HD video), and an intermediary application service, such as VoIP.
Each slice is designed to meet unique service assurance requirements and resource demands.
Fig. \ref{fig:e2e_network} demonstrates the implementation of our 5G network, which is reinforced by an isolated E2E network slice mechanism that leverages virtualization technology.
The 5G CN enables the construction of VNFs in a dynamic manner, as displayed by the User Plane Function (UPF) in our experimentation.
This enables a single network slice to accommodate either a singular VNF or a group of VNFs supported by a load balancer.
Hence, the CN has the capability of facilitating the scaling of VNFs in both the vertical and horizontal dimensions.
To balance traffic between VNF instances within a network slice, a weighted round-robin load balancing algorithm \cite{10.1007/978-981-13-5802-9_35} was integrated into the network slice manager.
In order to address the guaranteed bit rate requirements in network slicing, a Hierarchical Token Bucket (HTB) queue \cite{5272182} has been implemented in the router.
This queue has been shown to assist in isolating virtual network links in both TN and CN, thereby optimizing resources.
\begin{table}[h!t]
    \centering
    \begin{tabular}{ |p{1cm}||c|c|p{1cm}|p{1.2cm}| }
         \hline
         network slice & Transport type & Total & Scale factor & Network direction \\ [0.5ex] 
         \hline
         eMBB  & Cars                  & 9,075 & 1/25 & Downlink \\ 
         uRLLC & All trucks categories & 7,995 & 1/15 & Uplink \\
         mMTC  & Bikes and Motorcycles & 2,200 & 1/10 & Both \\
         VoIP  & Bus                   & 885 & 1/3 & Both\\
         \hline
        \end{tabular}
    \caption{Mapping from open data to the network slice devices.}
    \label{tab:map_vehicle_to_device}
\end{table}
\begin{figure*}[h]
    \centering
    \begin{subfigure}{0.95\columnwidth}
        \includegraphics[width=1\columnwidth]{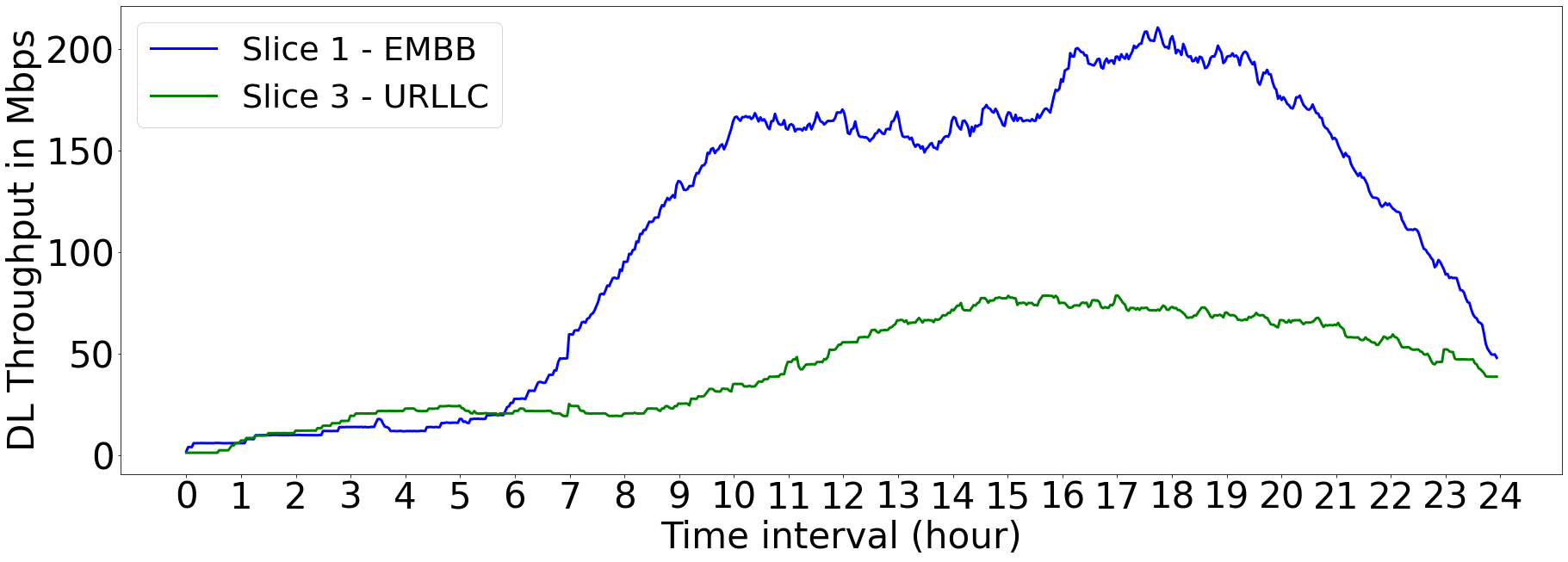}
        \caption{Downlink throughput eMBB and uRLLC slices}
    \end{subfigure}
    \hspace*{0.5cm}
    \begin{subfigure}{0.95\columnwidth}
        \includegraphics[width=1\columnwidth]{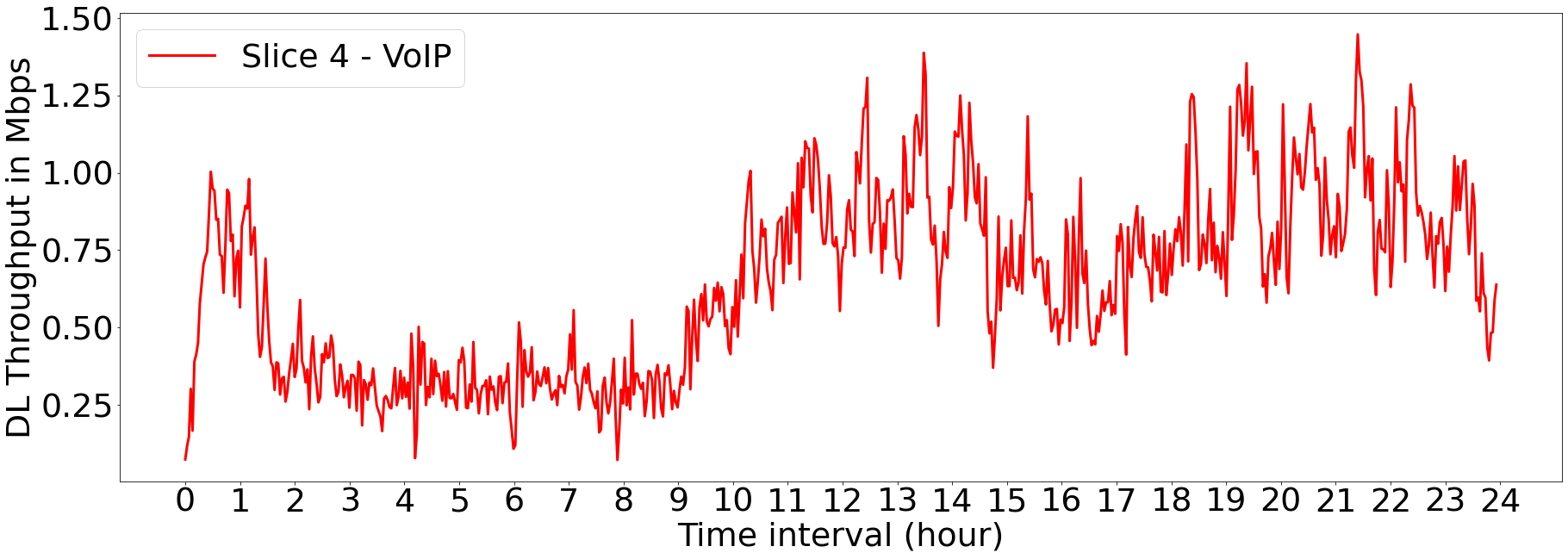}
        \caption{Downlink throughput VoIP slice}
    \end{subfigure}

    \begin{subfigure}{0.95\columnwidth}
        \includegraphics[width=1\columnwidth]{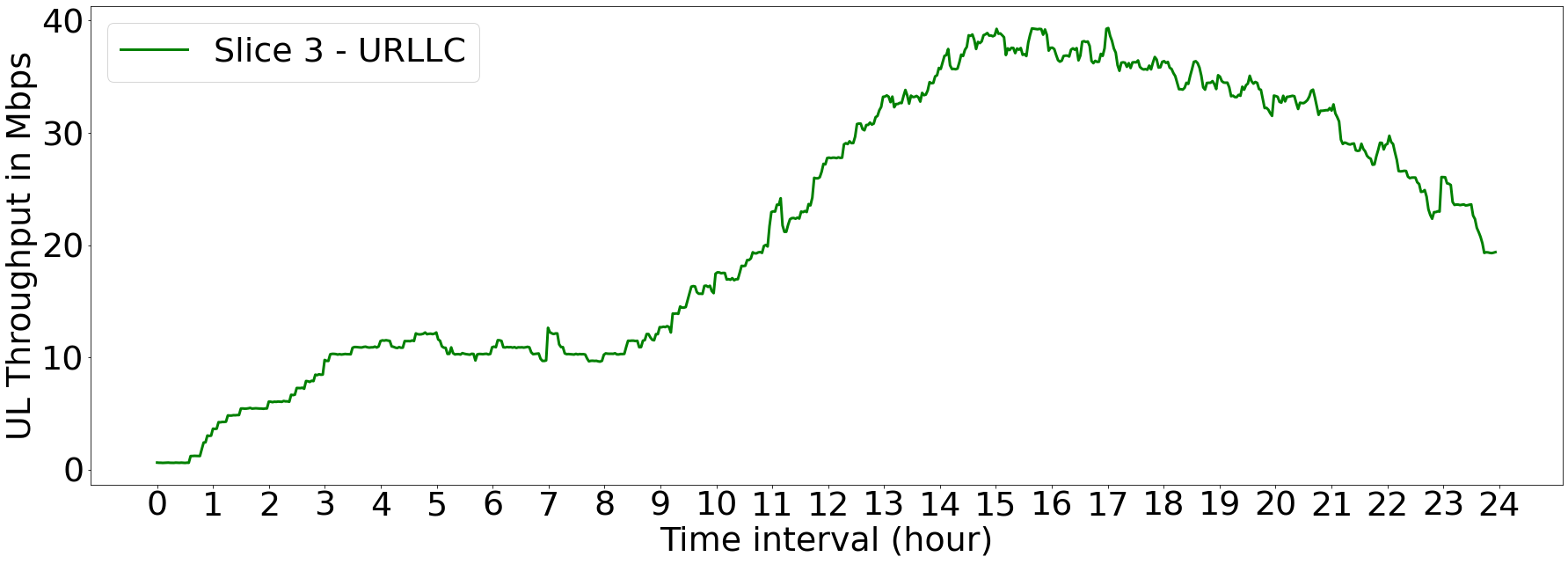}
        \caption{Uplink throughput uRLLC slice}
    \end{subfigure}
    \hspace*{0.5cm}
    \begin{subfigure}{0.95\columnwidth}
        \includegraphics[width=1\columnwidth]{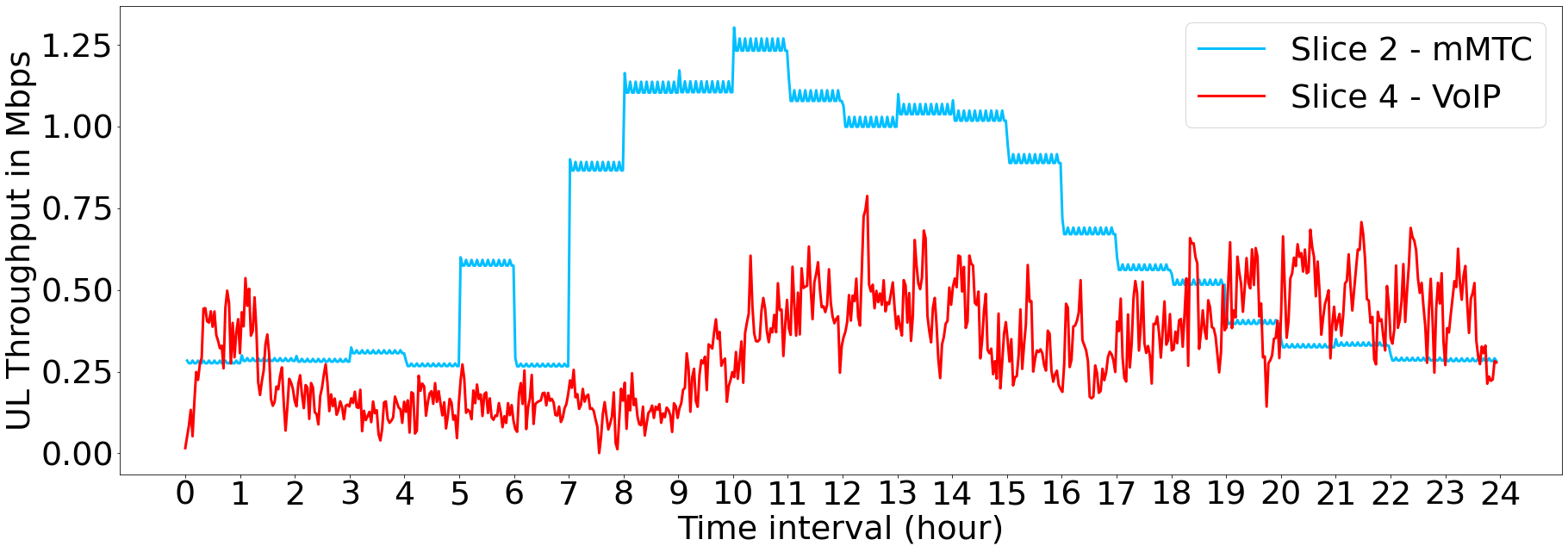}
        \caption{Uplink throughput mMTC and VoIP slices}
    \end{subfigure}

    \caption{The simulation of traffic patterns over a single day}
    \label{fig:simu_throughput}
\end{figure*}

Our simulation was configured to generate traffic patterns that closely resemble those found in real-world networks.
Table \ref{tab:map_vehicle_to_device} illustrates the specific number of UEs used in the four distinct network slices, with the UE types sourced from the open dataset \cite{opendatamtl}.
Using the data mentioned above, we compiled a summary of the number of mobile UEs present during each hour and randomly allocated their respective start positions within the 5G network.
Fig. \ref{fig:simu_throughput} depicts the testing environment, which is capable of accommodating diverse scenarios, including low and high peak traffic for both downlink and uplink directions (network slice 3 - mMTC), stable traffic (network slice 4 - VoIP), downlink direction (network slice 1 - eMBB) exclusively, and uplink direction (network slice 2 - uRLLC) exclusively.

\subsection{Evaluation of {\PSA}}

This section will evaluate {\PSA}'s performance in the service assurance domain within a 5G network environment.
The evaluation will be conducted within the simulation environment described above.
Our {\PSA} was designed to offer flexible configurations, aligning with infrastructure and network planning requirements.
It enables dynamic monitoring of network conditions, helping to identify potential resource issues and take appropriate actions to maintain high service quality for the network slice.
In the interest of simplicity and the capacity to readily discern the outcomes of our experiments, we employ uniform settings for all network slices. 
Nevertheless, it is feasible to configure disparate parameters for each network slice in practice.

{\PSA} was extensively examined to assess its effectiveness in two distinct layers: CN and TN layers.
The evaluation process included various factors, including latency, throughput, jitter, and packet loss, to maximize user experience.
To conduct a comprehensive analysis of the algorithm's capabilities, a set of E2E KPI limits and scale factors on different network slices was established and collected from different references \cite{5gAmeUrllc18, Siddiqi2019-we, 5GEVE2018-tt, 5gppp.5gpicture18, pop_2018}.
Detailed information about these KPI limits can be found in Table \ref{tab:kpiLimits}.
These factors play an important role in provisioning optimal service in 5G/B5G networks and provide insights into the algorithm's ability to detect anomalies, adapt to new network parameters, and make real-time adjustments to optimize service performance.

\begin{table}[ht]
\centering
\caption{End-to-end KPI limits and scale factors}
\label{tab:kpiLimits}
\resizebox{\columnwidth}{!}{%
\setlength{\tabcolsep}{2pt} 
\renewcommand{\arraystretch}{1.2} 
\begin{tabular}{|c|l||cccccccc|}
\hline

  \multirow{3}{*}{\begin{tabular}[c]{@{}c@{}}\textbf{KPI Type/} \\      \textbf{Unit}\end{tabular}} &
  \multicolumn{1}{c||}{\multirow{3}{*}{\textbf{KPI}}} &
  \multicolumn{8}{c|}{\textbf{Network slice type}} \\ \cline{3-10} 
 &
   &
  \multicolumn{2}{c|}{\textbf{eMBB}}  &
  \multicolumn{2}{c|}{\textbf{mMTC}} &
  \multicolumn{2}{c|}{\textbf{uRLLC}} &
  \multicolumn{2}{c|}{\textbf{VoIP}}
  \\ \cline{3-10} 
 &
   &
  \multicolumn{1}{c|}{\textbf{Th}} &
  \multicolumn{1}{c|}{\textbf{SF}} &
  \multicolumn{1}{c|}{\textbf{Th}} &
  \multicolumn{1}{c|}{\textbf{SF}} &
  \multicolumn{1}{c|}{\textbf{Th}} &
  \multicolumn{1}{c|}{\textbf{SF}} &
 \multicolumn{1}{c|}{\textbf{Th}} &
  \multicolumn{1}{c|}{\textbf{SF}}  \\ 
  \hline
  \hline

  \multirow{2}{*}{\begin{tabular}[c]{@{}c@{}}Delay\\      ms\end{tabular}} &
  Average Delay &
  \multicolumn{1}{c|}{\multirow{2}{*}{300}} &
  \multicolumn{1}{c|}{\multirow{2}{*}{0.20}} &
  \multicolumn{1}{c|}{\multirow{2}{*}{10}} &
  \multicolumn{1}{c|}{\multirow{2}{*}{2.5}} &
  \multicolumn{1}{c|}{\multirow{2}{*}{30$_{(i)}$}} &
  \multicolumn{1}{c|}{\multirow{2}{*}{2}} &
  \multicolumn{1}{c|}{\multirow{2}{*}{100}} &
  \multirow{2}{*}{0.25} \\ \cline{2-2}
 &
  Max Delay &
  \multicolumn{1}{c|}{} &
  \multicolumn{1}{c|}{} &
  \multicolumn{1}{c|}{} &
  \multicolumn{1}{c|}{} &
  \multicolumn{1}{c|}{} &
  \multicolumn{1}{c|}{} &
  \multicolumn{1}{c|}{} &
   \\ \hline

  \multirow{2}{*}{\begin{tabular}[c]{@{}c@{}}Jitter\\      ms\end{tabular}} &
  \multirow{2}{*}{Jitter} &
  \multicolumn{1}{c|}{\multirow{2}{*}{100}} &
  \multicolumn{1}{c|}{\multirow{2}{*}{0.012}} &
  \multicolumn{1}{c|}{\multirow{2}{*}{N/A$_{(iv)}$}} &
  \multicolumn{1}{c|}{\multirow{2}{*}{N/A}} &
  \multicolumn{1}{c|}{\multirow{2}{*}{5$_{(iii)}$}} &
  \multicolumn{1}{c|}{\multirow{2}{*}{\begin{tabular}[c]{@{}c@{}}1.05 (UL)/\\      1 (DL)\end{tabular}}} &
  \multicolumn{1}{c|}{\multirow{2}{*}{10$_{(iv)}$}} &
  \multirow{2}{*}{\begin{tabular}[c]{@{}c@{}}0.2 (UL)/\\      06 (DL)\end{tabular}} \\
 &
   &
  \multicolumn{1}{c|}{} &
  \multicolumn{1}{c|}{} &
  \multicolumn{1}{c|}{} &
  \multicolumn{1}{c|}{} &
  \multicolumn{1}{c|}{} &
  \multicolumn{1}{c|}{} &
  \multicolumn{1}{c|}{} &
   \\ \hline

  \begin{tabular}[c]{@{}c@{}}Packet   loss\\      \%\end{tabular} &
  Packet loss &
  \multicolumn{1}{c|}{1E-03$_{(v)}$} &
  \multicolumn{1}{c|}{1E+03} &
  \multicolumn{1}{c|}{1E-02} &
  \multicolumn{1}{c|}{285} &
  \multicolumn{1}{c|}{0.1$_{(ii)}$} &
  \multicolumn{1}{c|}{10} &
  \multicolumn{1}{c|}{1.00} &
  2.5 \\ \hline

  \multirow{2}{*}{\begin{tabular}[c]{@{}c@{}}Throughput \\      Kbps\end{tabular}} &
  \multirow{2}{*}{Throughput} &
  \multicolumn{1}{c|}{\multirow{2}{*}{N/A}} &
  \multicolumn{1}{c|}{\multirow{2}{*}{N/A}} &
  \multicolumn{1}{c|}{\multirow{2}{*}{N/A}} &
  \multicolumn{1}{c|}{\multirow{2}{*}{N/A}} &
  \multicolumn{1}{c|}{\multirow{2}{*}{N/A}} &
  \multicolumn{1}{c|}{\multirow{2}{*}{N/A}} &
  \multicolumn{1}{c|}{\multirow{2}{*}{N/A}} &
  \multirow{2}{*}{N/A} \\
 &
   &
  \multicolumn{1}{c|}{} &
  \multicolumn{1}{c|}{} &
  \multicolumn{1}{c|}{} &
  \multicolumn{1}{c|}{} &
  \multicolumn{1}{c|}{} &
  \multicolumn{1}{c|}{} &
  \multicolumn{1}{c|}{} &
   \\ \hline
\end{tabular}
}

 {\raggedright 
    \vspace{1ex}
 \scriptsize 
 \textbf{Th}: Threshold; \textbf{CF}: Scale Factor; 
 \textit{Source}: ($i$) Table 3.1 \cite{5gAmeUrllc18};
 ($ii$) Table 3 \cite{Siddiqi2019-we};
 ($iii$) Tables 14 \cite{5GEVE2018-tt};
 ($iv$) Table 3\&10 \cite{5gppp.5gpicture18};
 ($v$) Section I \cite{pop_2018}
 }
\end{table}

\begin{table}[ht]
    \centering
    \begin{tabular}{|c|l|l|l|}
        \hline
        Notation & Setting 1 & Setting 2 & Setting 3\\
        \hline
        $\ratioOverProvision$ & .15 & \color{red}{.1} & \color{red}{.05} \\
        $\ratioScaling$ &  .05 &  - & - \\
        $\TrafficModelAccuary$ & .814 & - & - \\
        $\ratioResUsage$ & .8 & - & - \\
        $\ratioValidateScaling$ & .02 & - & - \\
        $\MaxCAPVNFSlice$ & 3 CPUs, 1 GB, 1.2 GB & - & - \\
        $\MinCAPVNFSlice$ & .1 CPU, 15 MB, 20 MB & - & - \\
        $\NumberOfSampling$   & 15 minutes samples & - & - \\
        $\totalPhyCap$        & 39 CPUs, 13 GB, 15 GB & - & - \\
        $\totalLinkCap$ & 500 Mbps & - & - \\
        $\TimeWindow$ & 5 mins & - & - \\
        $\MonitorTrafficTime$ & 2 $\cdot \TimeWindow$ & - & - \\
        \hline
    \end{tabular} \\
        \vspace{1ex}
        {\raggedright -: same as setting 1. \par}
    \caption{Evaluate algorithm parameters (not including KPIs).}
    \label{tab:evaluation_parameters}
\end{table}

\begin{table*}[ht]
    \centering
    \begin{tabular}{|c|c|c|c|c|c|c|c|c|c|}
    \hline
        Setting & \makecell{Network\\slice} & \makecell{Number of \\KPI violation} & \makecell{Total\\action} & Scale\_down & Scale\_up & Scale\_out & Scale\_in & \makecell{Ratio action \\/ simulation time}  & \makecell{Ratio action \\hourly}\\
        \hline
        1 & \makecell{eMBB \\ uRLLC \\ mMTC \\ VoIP} & \makecell{0 \\0 \\0 \\0} & \makecell{238 \\130 \\223 \\136} & \makecell{15 \\29 \\8 \\20} & \makecell{216 \\101 \\211 \\116} & \makecell{4 \\0 \\2 \\0} & \makecell{3 \\0 \\2 \\0} & \makecell{33\% \\18.2\% \\30\% \\19.7\%} & \makecell{1.38\% \\.76\% \\1.25\% \\.86\%}  \\
        \hline
        2 & \makecell{eMBB \\ uRLLC \\ mMTC \\ VoIP} & \makecell{9 (delays) \\0 \\0 \\0} & \makecell{243 \\154 \\65 \\245} & \makecell{13 \\33 \\20 \\63} & \makecell{219 \\ 121 \\232 \\182} & \makecell{6 \\0 \\7 \\0} & \makecell{5 \\0 \\6 \\0}  & \makecell{33.9\% \\21.5\% \\37\% \\35.3\%} & \makecell{1.41\% \\.9\% \\1.54\% \\1.47\%}  \\
        \hline
        3 & \makecell{eMBB \\ uRLLC \\ mMTC \\ VoIP} & \makecell{24 (delays), 2 (packet losses) \\0 \\1 (delay) \\0} & \makecell{349 \\168 \\321 \\294} & \makecell{104 \\45 \\50 \\101} & \makecell{234 \\ 123 \\258 \\193} & \makecell{6 \\0 \\7 \\0} & \makecell{5 \\0 \\6 \\0}  & \makecell{48.7\% \\23.5\% \\44.8\% \\42.3\%} & \makecell{2.03\% \\.99\% \\1.86\% \\1.76\%}  \\
        \hline
    \end{tabular}
    \caption{Summary of the results obtained from the algorithm over 24 hours with different settings}
    \label{tab:summary_algorithm_results}
\end{table*}

\begin{figure}[h]
    \centering
    \begin{subfigure}{1\columnwidth}
        \includegraphics[width=1\columnwidth]{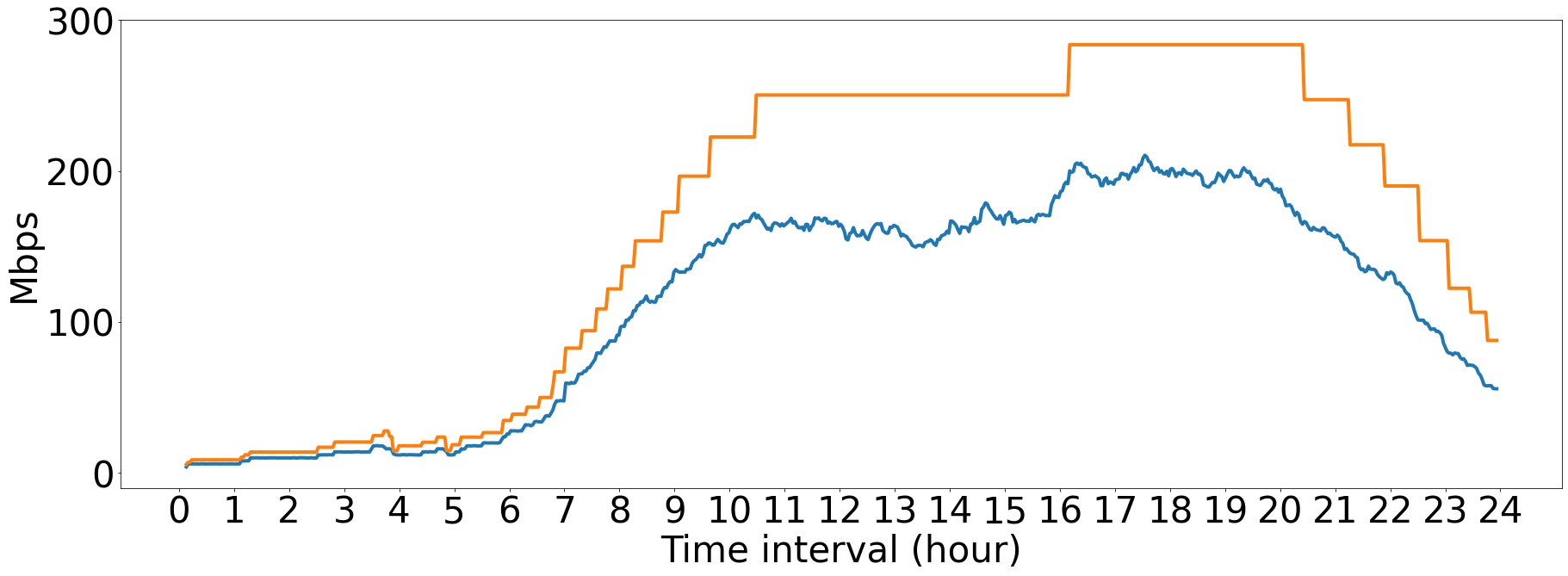}
        \caption{Virtual link capacity configuration for eMBB slice at TN layer (Orange: Configuration provided by closed-loop algorithm, Blue: actual network throughput)}
    \end{subfigure}

    \begin{subfigure}{1\columnwidth}
        \includegraphics[width=1\columnwidth]{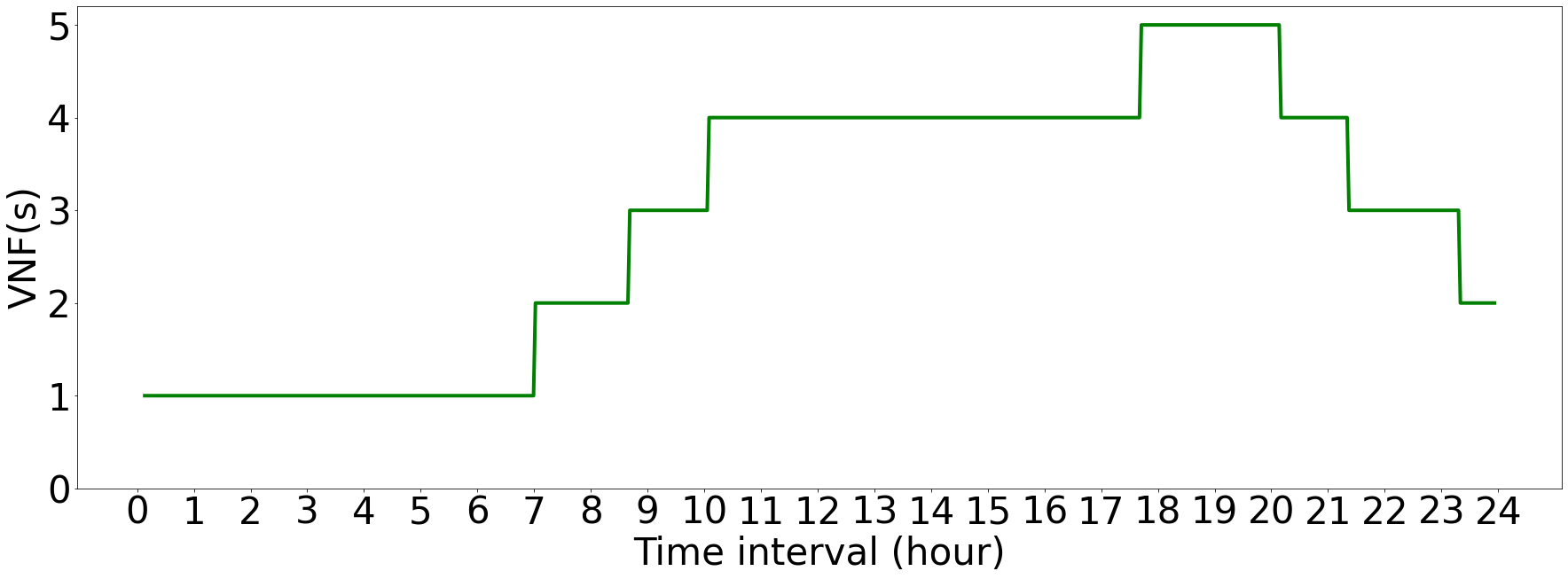}
        \caption{Number of VNFs utilized in eMBB slice}
    \end{subfigure}
    \hspace*{1.cm}
    \caption{Virtual link capacity configuration and number of VNFs assigned to eMBB slice - Setting 1}
    \label{fig:s1_traffic_boundary}
\end{figure}

\begin{figure*}[h]
    \centering
    \begin{subfigure}{1.1\columnwidth}
        \includegraphics[width=1\columnwidth]{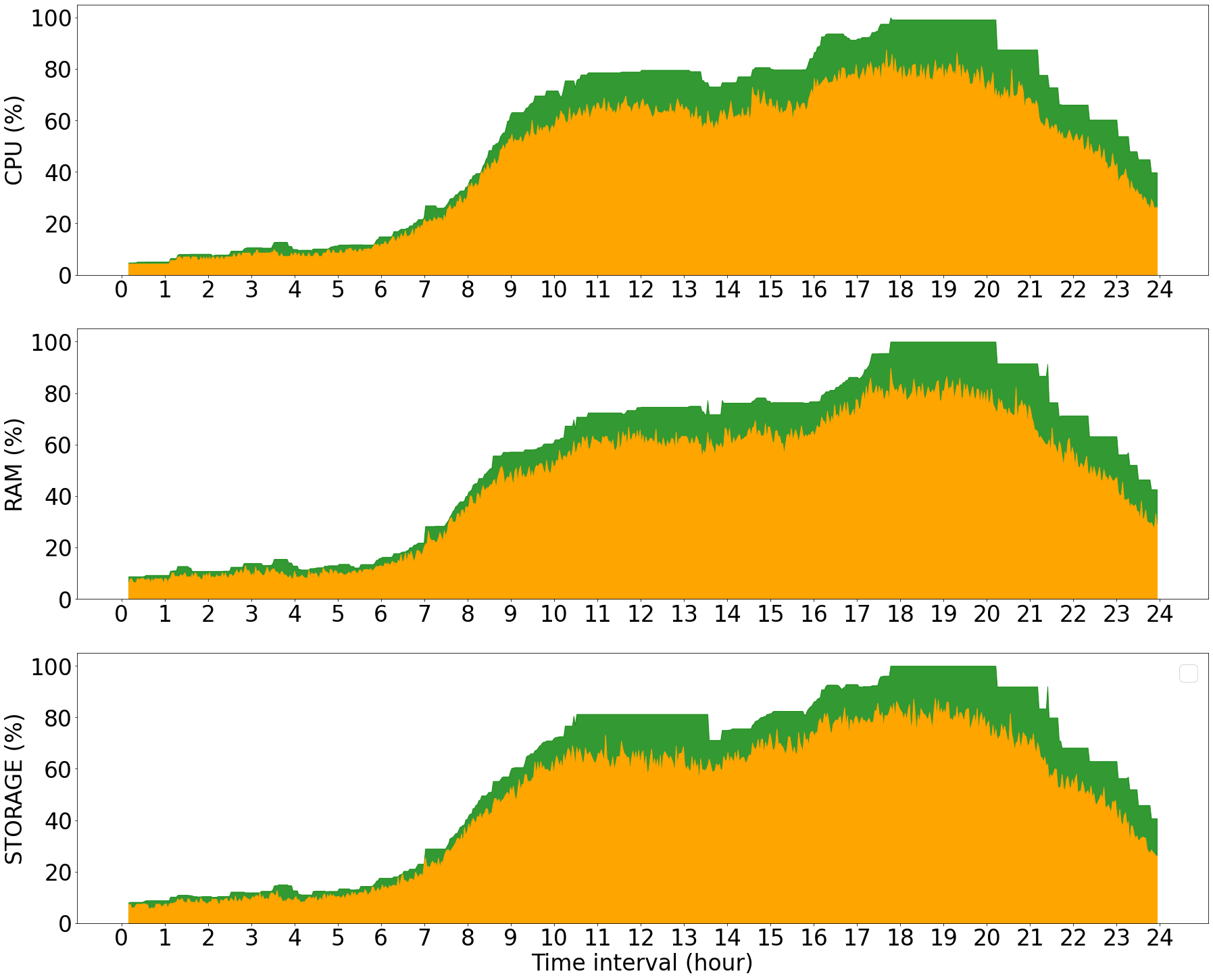}
        \caption{Spare resources (green) and resources utilization (orange)}
    \end{subfigure}
    \begin{subfigure}{0.65\columnwidth}
        \begin{subfigure}{1\columnwidth}
            \includegraphics[width=1.1\columnwidth]{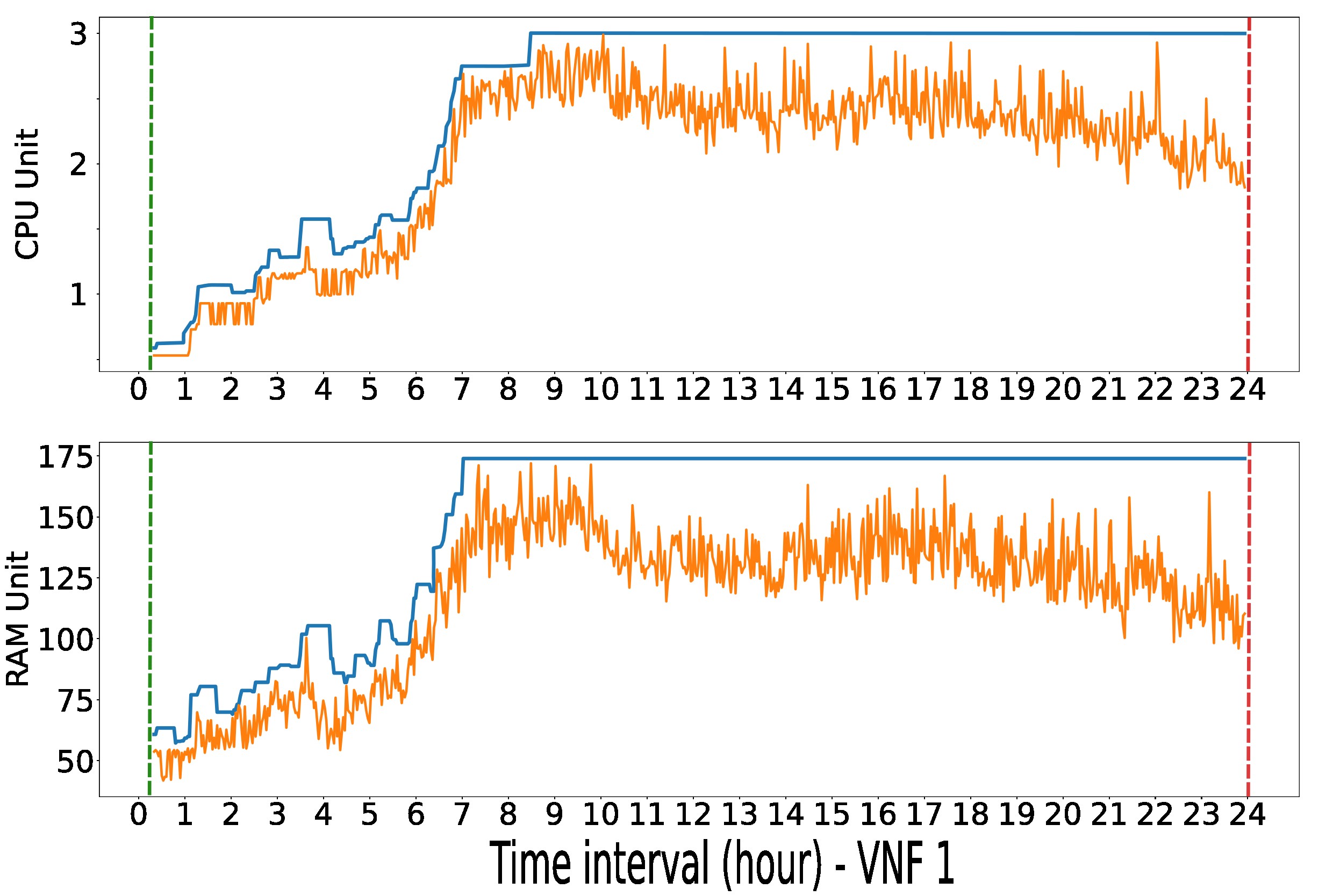}
            \caption{VNF 1}
        \end{subfigure}
        \begin{subfigure}{1\columnwidth}
            \includegraphics[width=1.1\columnwidth]{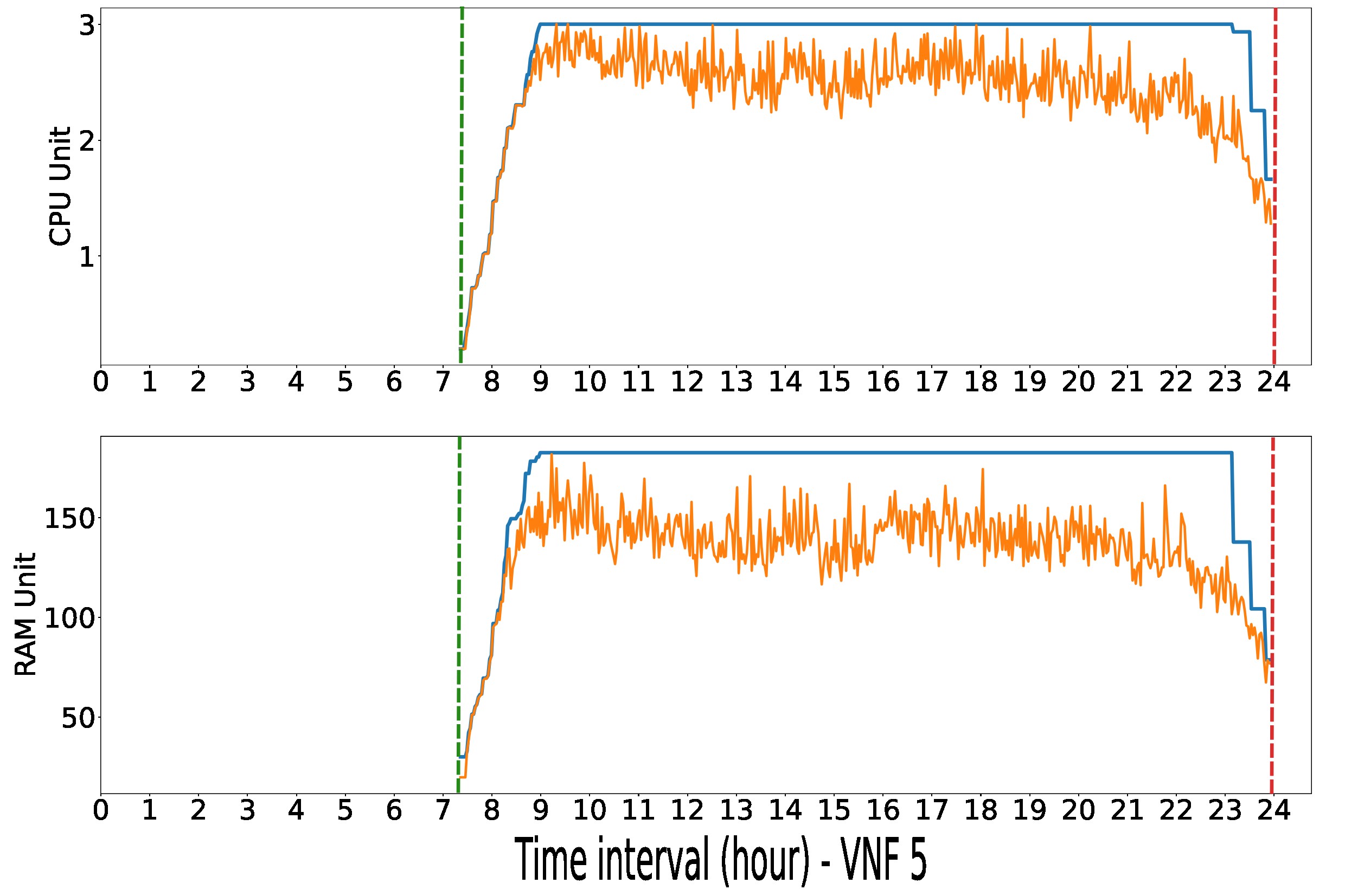}
            \caption{VNF 5}
        \end{subfigure}
    \end{subfigure}
    
    \begin{subfigure}{.66\columnwidth}
        \includegraphics[width=1\columnwidth]{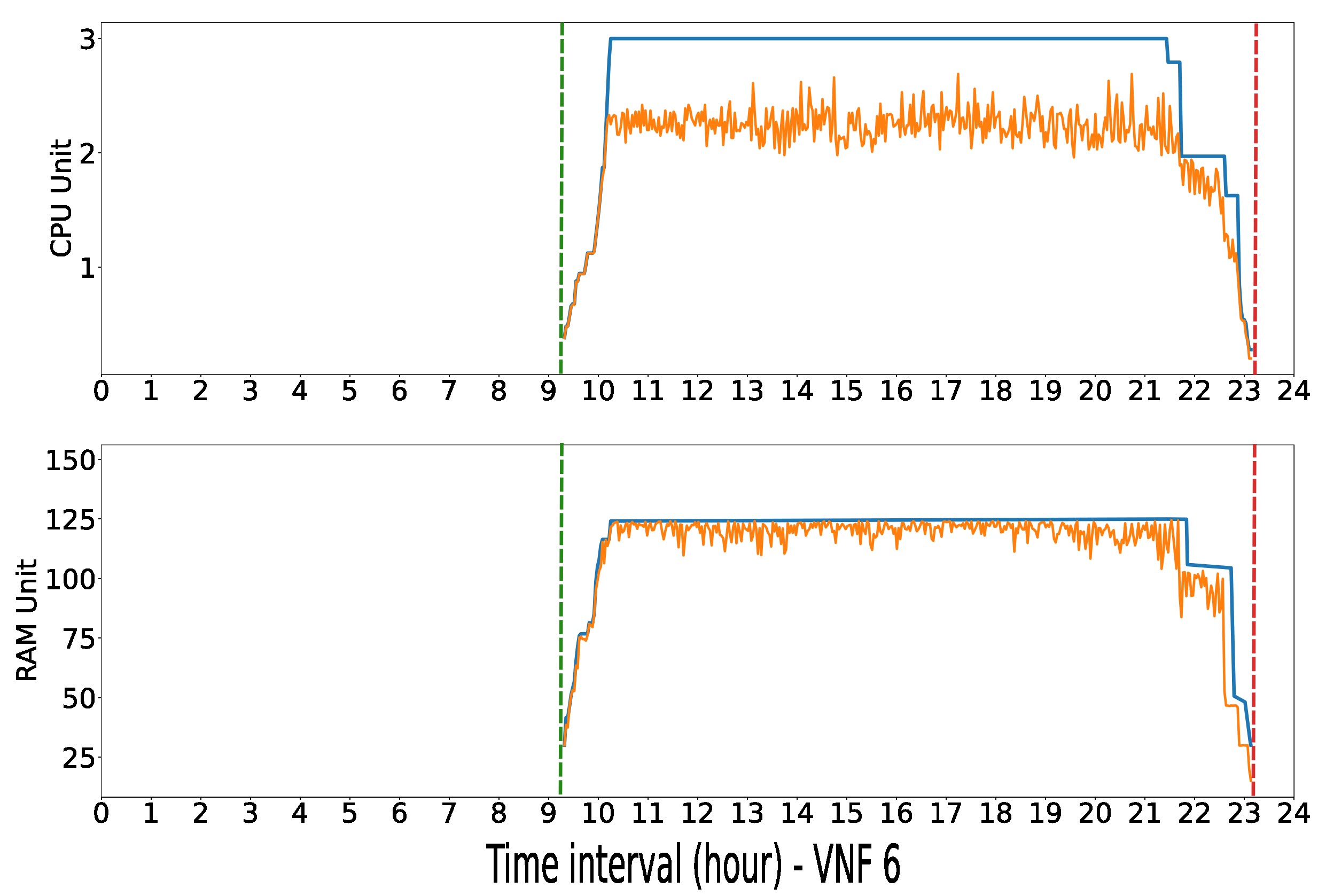}
        \caption{VNF 6}
    \end{subfigure}
    \hspace*{0.001cm}
    \begin{subfigure}{.66\columnwidth}
        \includegraphics[width=1\columnwidth]{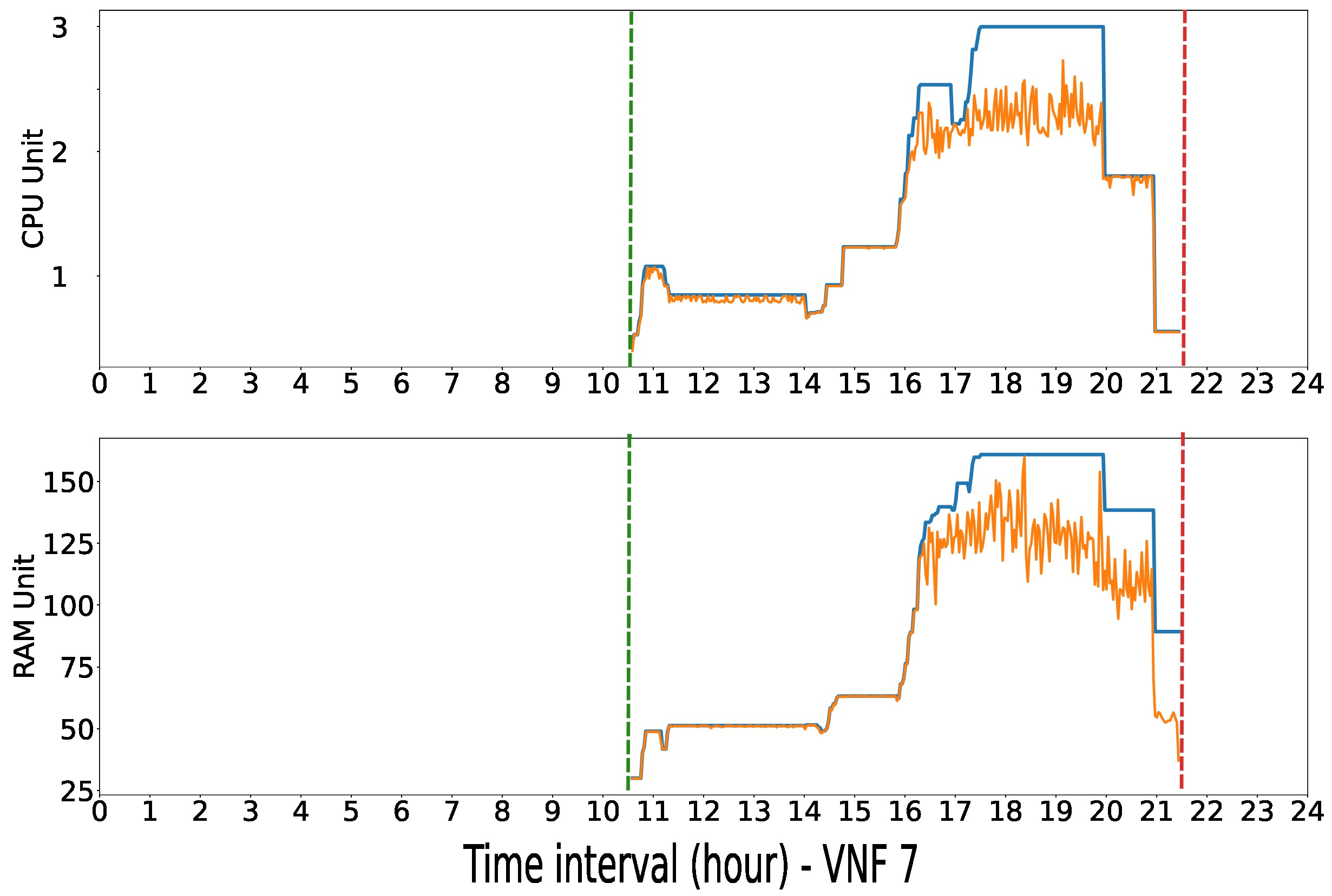}
        \caption{VNF 7}
    \end{subfigure}
    \hspace*{0.001cm}
    \begin{subfigure}{.66\columnwidth}
        \includegraphics[width=1\columnwidth]{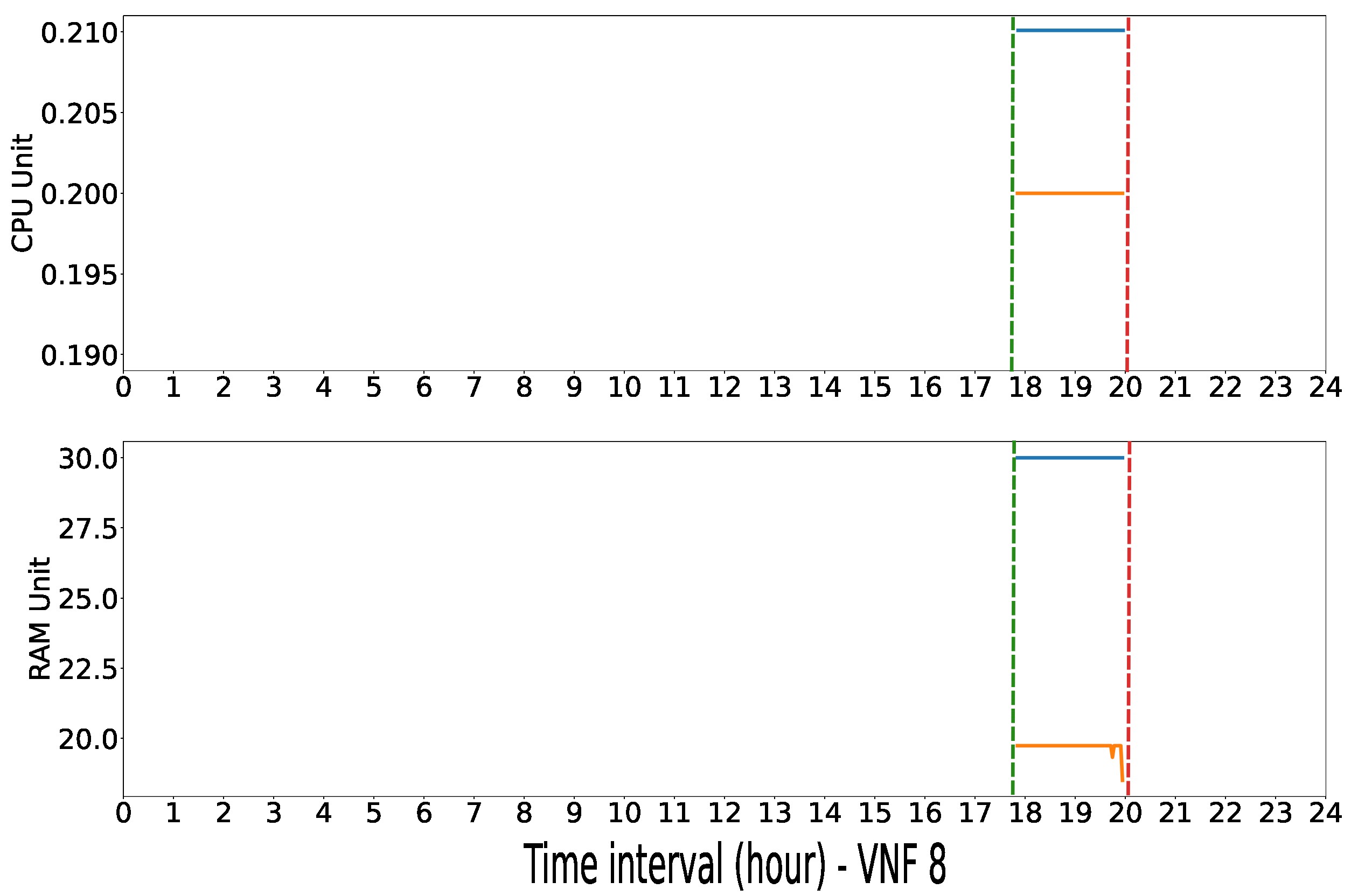}
        \caption{VNF 8}
    \end{subfigure}

    \caption{The utilization of VNFs allocated to the eMBB slice with setting 1. (vertical green line: time of adding VNF, vertical red line: time of removing VNF, orange line: actual resource utilization, blue line: configured resource.)}
    \label{fig:s1_resources_utilization}
\end{figure*}

The configuration parameters utilized in the evaluation environment are delineated in Table \ref{tab:evaluation_parameters}.
During the evaluation phase, the algorithm successfully determined the data rate needed to be used as a traffic boundary for each network slice, even when the network slice exhibited a significantly high data rate.
By leveraging an advanced ML traffic forecasting model, {\PSA} reliably guarantees the bit rate and ensures a seamless flow of traffic.
For a more detailed view of the output configuration, refer to Fig. \ref{fig:s1_traffic_boundary} (a) as an example of the eMBB slice.
Meanwhile, the algorithm demonstrates its proficiency in optimizing the allocation of resources for the VNF instances, as depicted in Fig. \ref{fig:s1_resources_utilization}.
In the figure, the orange colour represents the actual resource utilization, while the blue colour indicates the configured resources for the VNF instances.
It is evident that the algorithm excels in predicting resource utilization and proactively allocating resources accordingly.
As illustrated in Fig. \ref{fig:s1_resources_utilization}(a) for the overall utilization of all VNF instances of the eMBB slice, the algorithm closely provides optimal resources for the network slice and always allocates spare resources in advance in accordance with the requirements of the slice.
To elaborate on Fig. \ref{fig:s1_resources_utilization}(b to f), the algorithm is capable of providing the necessary resources for each VNF instance while being able to dynamically add or remove instances correctly to optimize resources in accordance with the requirements of the slice.
Therefore, this indicates that the algorithm offers the capability to identify and allocate resources in an optimized way.
Nevertheless, it is important to note that the distribution of spare resources and the execution of actions of the algorithm might vary based on the parameters $\ratioOverProvision, \ratioResUsage, \ratioValidateScaling $ and $\ratioScaling$.
If their values are sufficiently small, the algorithm will probably execute actions more frequently, as the available resources will be depleted sooner, but the network will save more resources.

\begin{figure*}[h]
    \centering

    \begin{subfigure}{0.95\columnwidth}
        \includegraphics[width=1\columnwidth]{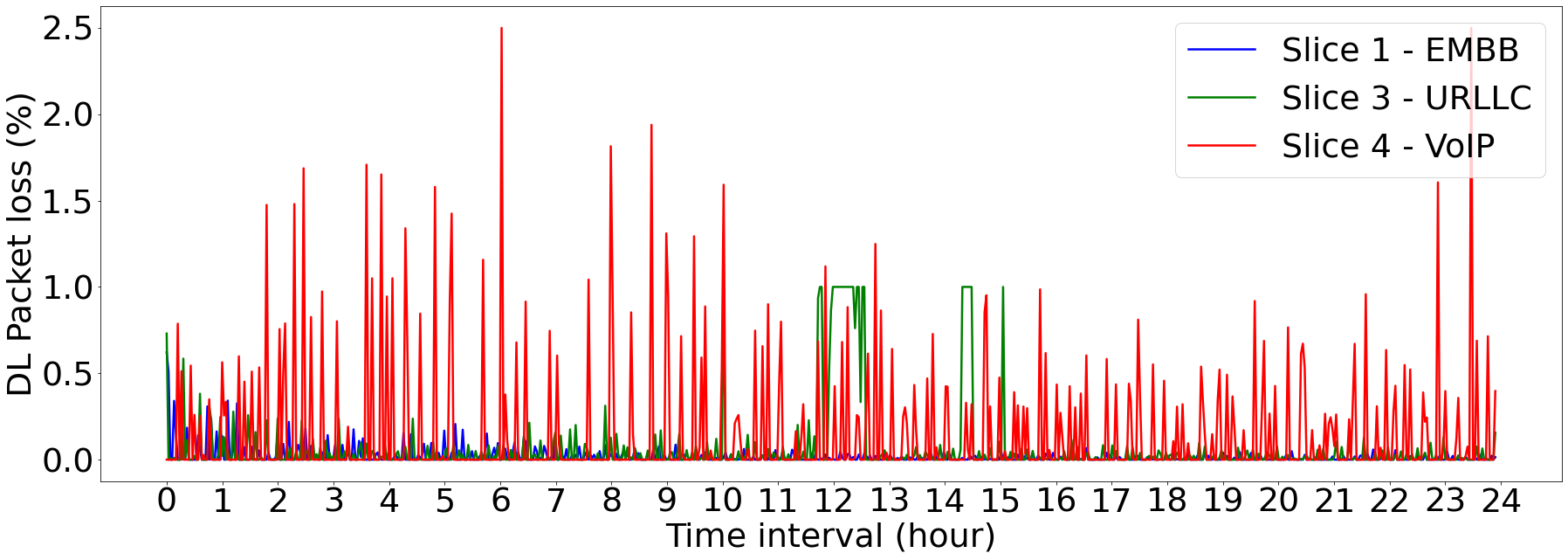}
        \caption{Downlink packet loss}
    \end{subfigure}
    \hspace*{0.5cm}
    \begin{subfigure}{0.95\columnwidth}
        \includegraphics[width=1\columnwidth]{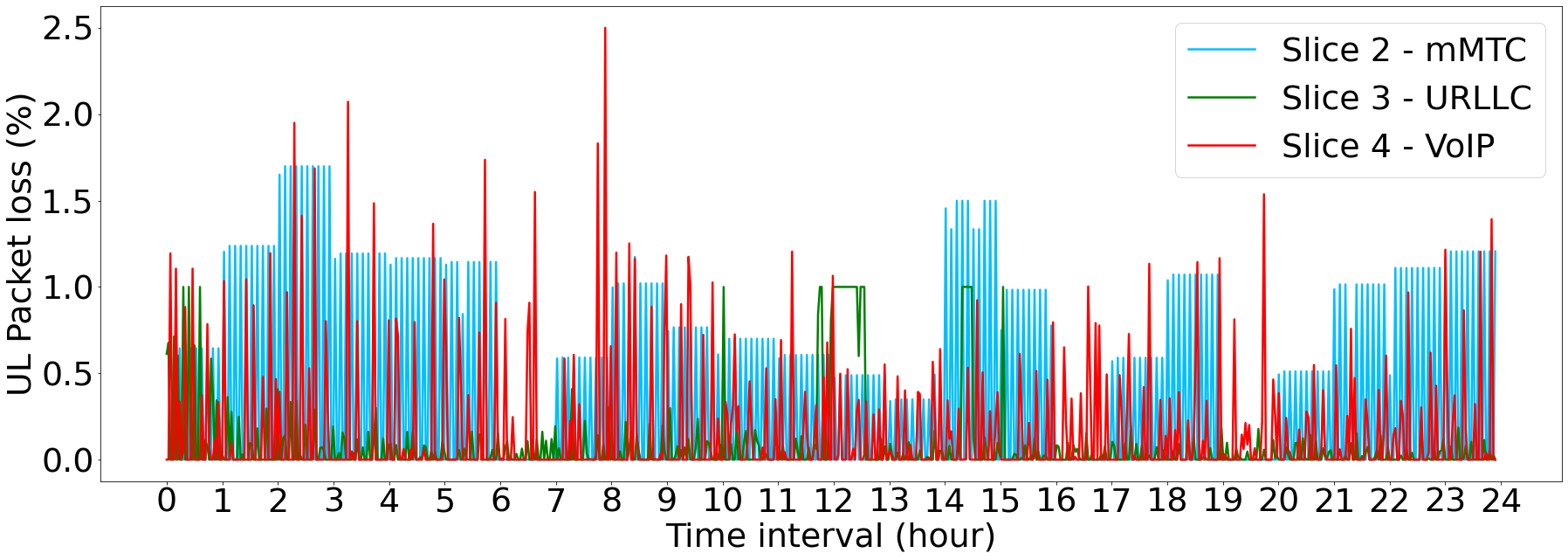}
        \caption{Uplink packet loss}
    \end{subfigure}

    \begin{subfigure}{0.95\columnwidth}
        \includegraphics[width=1\columnwidth]{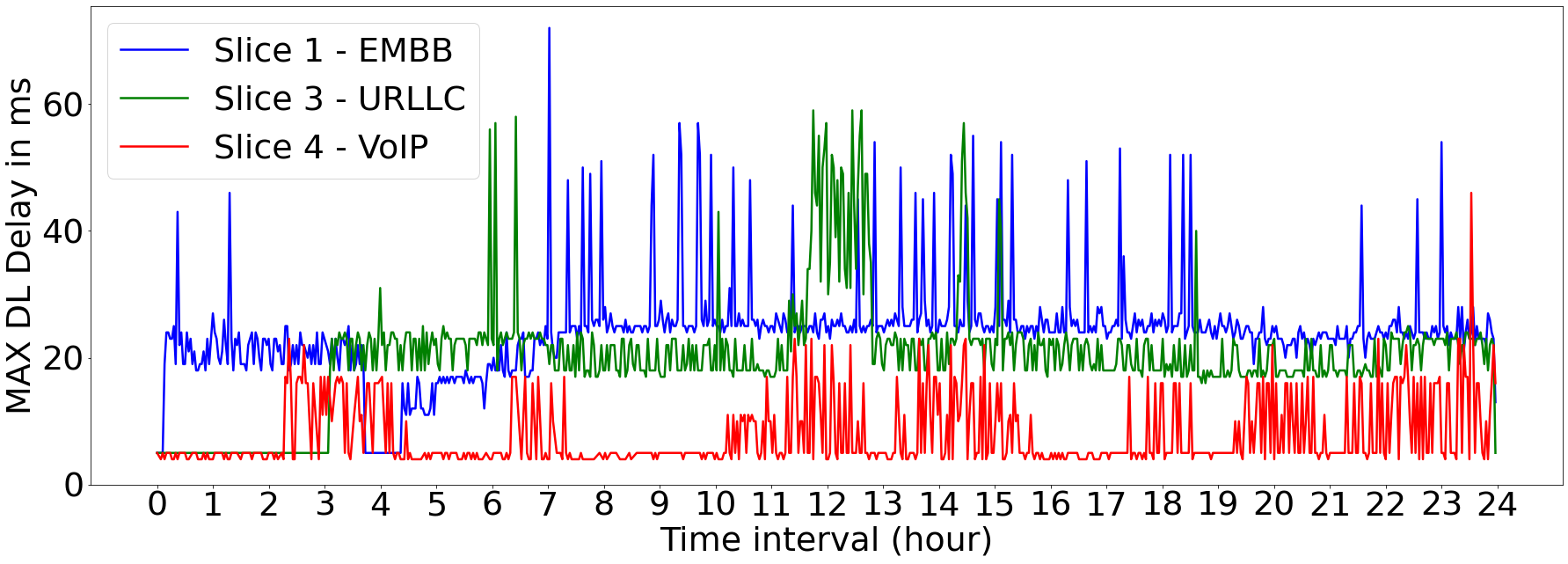}
        \caption{Downlink delay}
    \end{subfigure}
    \hspace*{0.5cm}
    \begin{subfigure}{0.95\columnwidth}
        \includegraphics[width=1\columnwidth]{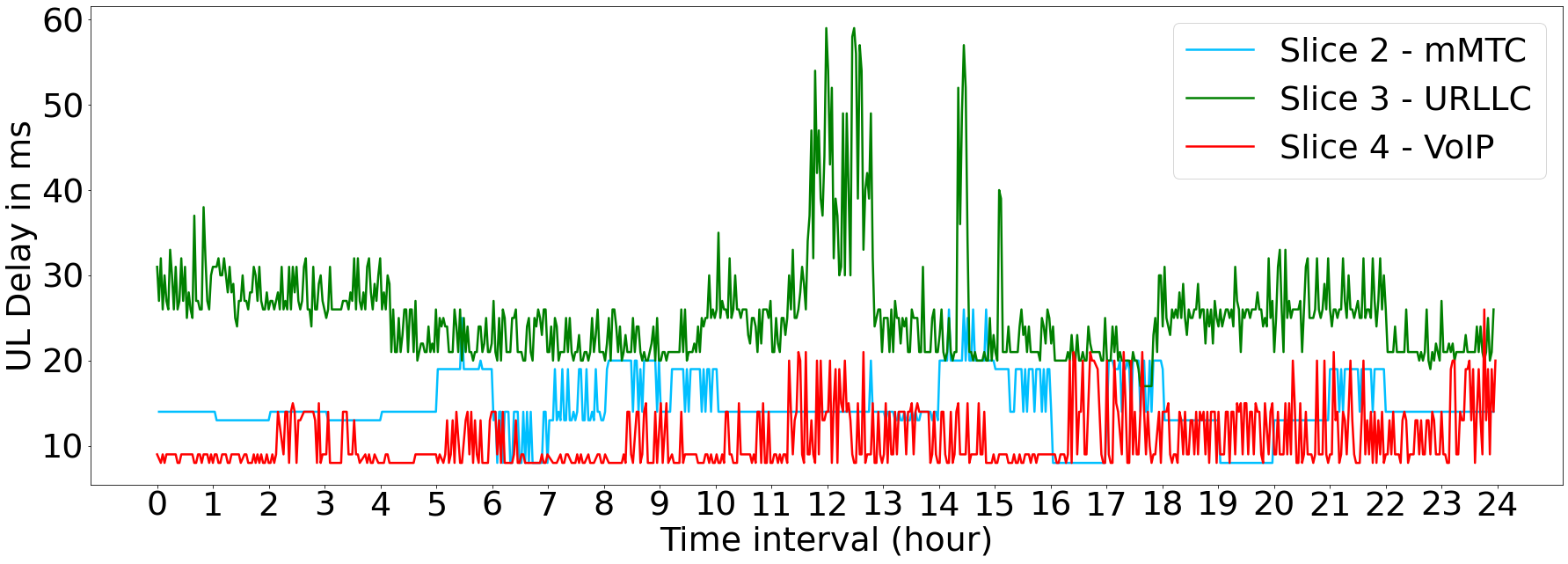}
        \caption{Uplink delay}
    \end{subfigure}

    \begin{subfigure}{0.95\columnwidth}
        \includegraphics[width=1\columnwidth]{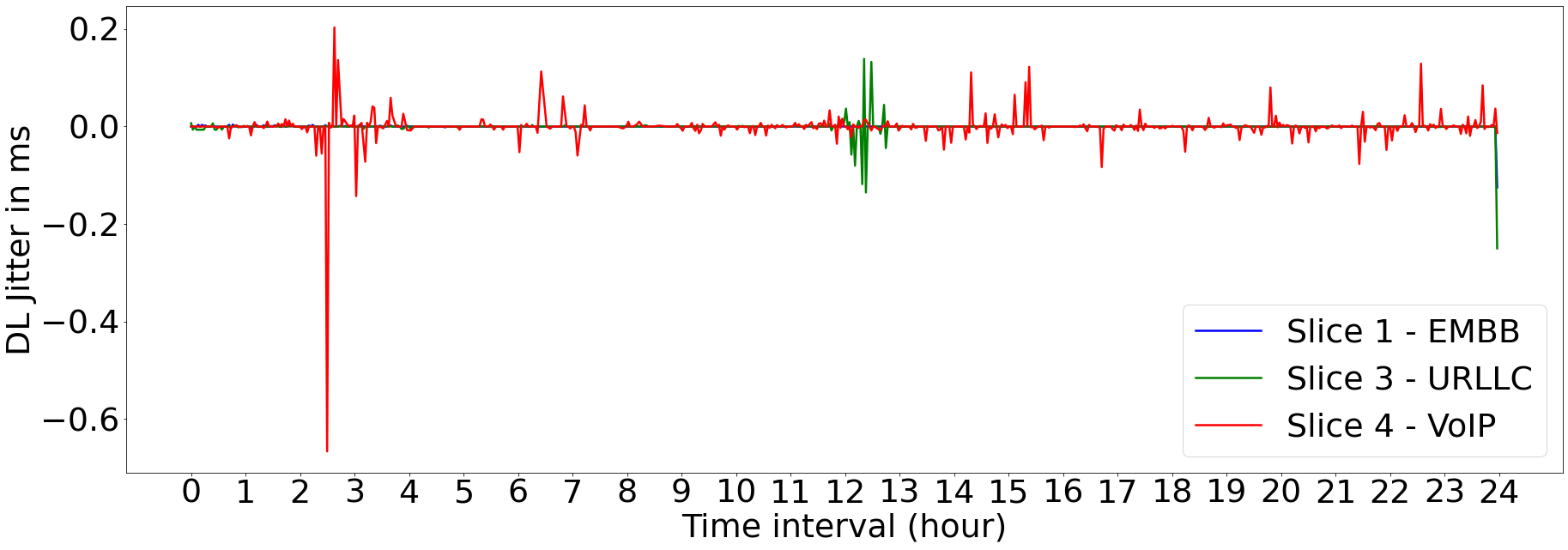}
        \caption{Downlink jitter}
    \end{subfigure}
    \hspace*{0.5cm}
    \begin{subfigure}{0.95\columnwidth}
        \includegraphics[width=1\columnwidth]{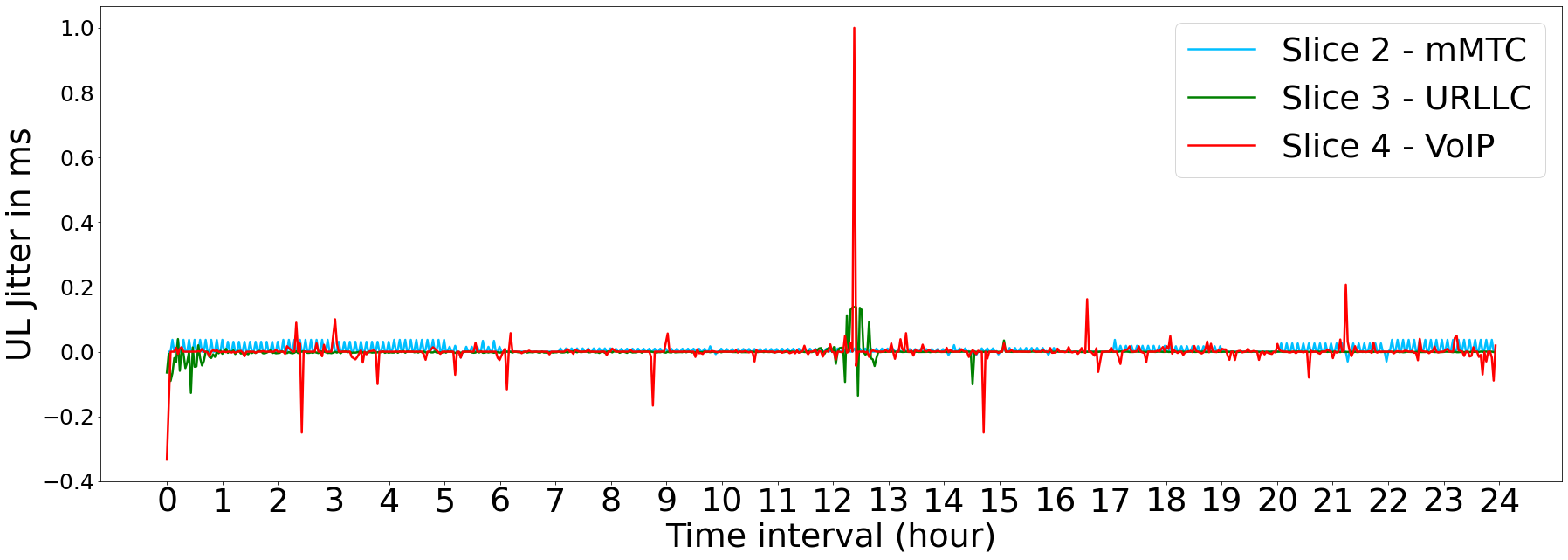}
        \caption{Uplink jitter}
    \end{subfigure}

    \caption{End-to-end network KPIs with the support of proactive closed-loop algorithm - Setting 1}
    \label{fig:end_to_end_kpi}
\end{figure*}

A comprehensive overview of the {\PSA} performance across four network slices is provided in both Table \ref{tab:summary_algorithm_results} and Fig. \ref{fig:end_to_end_kpi}.
In detail, {\PSA} proves effective at minimizing KPI violations across various settings, as presented in Table \ref{tab:evaluation_parameters}. 
This is achieved even during high traffic spikes and network condition changes, as demonstrated in the use cases of the eMBB and uRLLC slices.
Furthermore, it strikes a balance between the number of actions taken (ranging from 1\% to 2\% on an hourly basis) and the allocation of spare resources to the network slices.
As depicted in Fig. \ref{fig:end_to_end_kpi}, the algorithm with setting 1 successfully prevents KPI violations across all network slices, in terms of packet loss (a, b), delay (c, d), and jitter (e, f), thus meeting our QoS targets.
In addition, the results also demonstrate the efficacy of {\PSA} in executing parallel operations with a high level of performance.
It quickly and accurately identifies KPI violations, performs corrective actions, and allocates appropriate resources to resolve issues promptly.
Therefore, the {\PSA} effectively reduces the number of KPI violations over time, leading to improved overall QoS for the network and enhanced QoE for end users.

\begin{figure}[h]
    \centering
    \includegraphics[width=0.92\columnwidth]{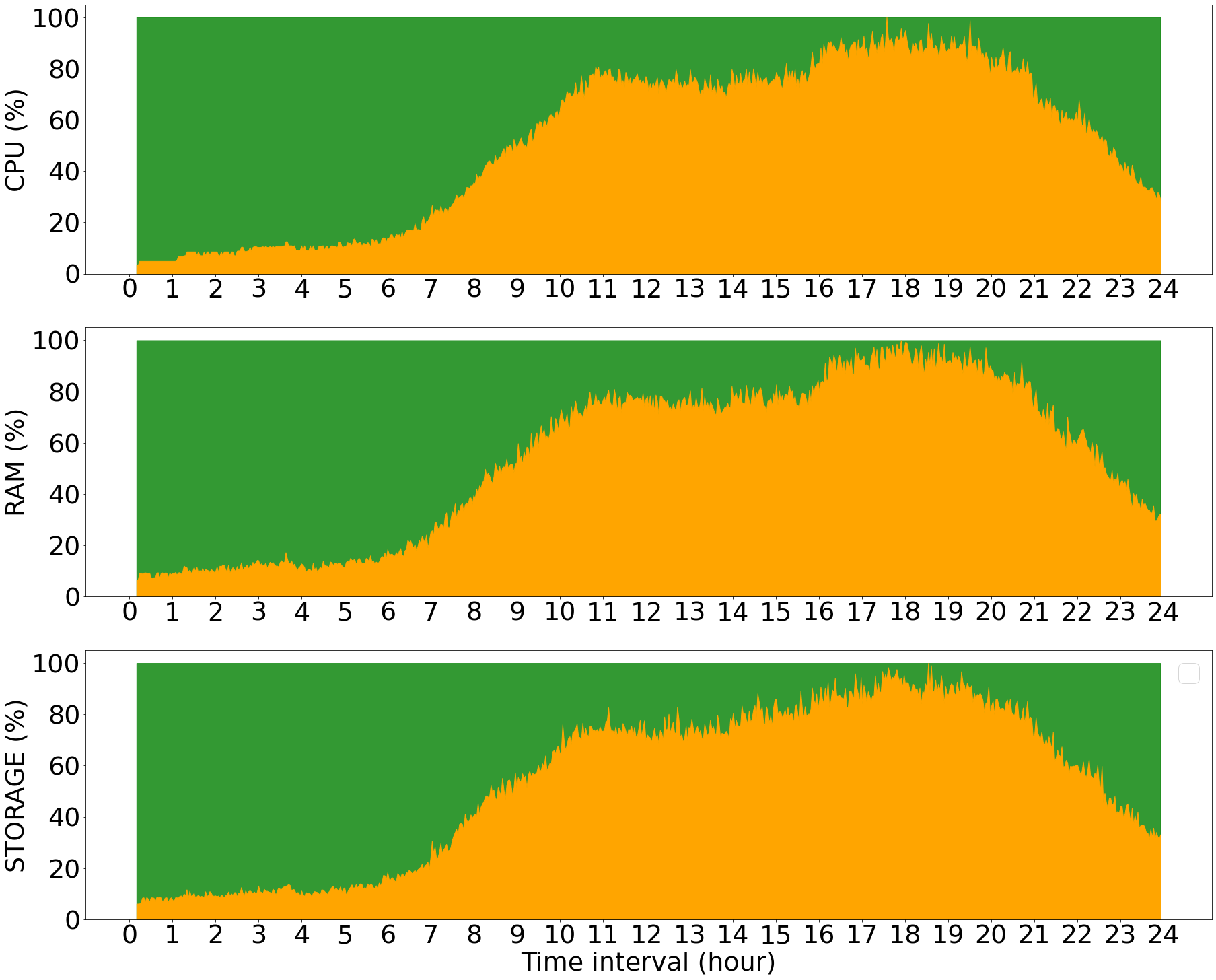}
    \caption{Resource allocation without our closed-loop algorithm in the eMBB slice during the whole simulation}
    \label{fig:spare_res_optimal_withoucl}
\end{figure}

Assuming that the highest peak of network traffic is known and sufficient resources are configured for a network slice to operate efficiently without any KPI violation.
As illustrated in Figure \ref{fig:spare_res_optimal_withoucl}, spare resources are represented in green, while actual eMBB resource consumption is depicted in orange.
In comparison to this worst-case scenario, our {\PSA} can reduce resource consumption by \textbf{54.85\%} for the eMBB slice (see Fig. \ref{fig:s1_resources_utilization}(a) and Fig. \ref{fig:spare_res_optimal_withoucl}).
The overall resource savings are calculated as the mean difference between the total resources used with and without the algorithm enabled, over the entire simulation period.
Across disparate network slices, the algorithm has been shown to achieve significant aggregate resource savings. Specifically, resource savings of 50.87\% for mMTC, 57.1\% for uRLLC, and 23.63\% for VoIP were observed.
The relatively lower savings for VoIP can be attributed to its stable traffic, as discussed in Section \ref{sec:experiment}-A.
Additionally, we observed an inverse correlation between the accepted over-provisioning rate and the number of scaling actions.
As the over-provisioning rate increases, the number of scaling actions decreases.
This is due to the trade-off between resource utilization and service assurance: higher over-provisioning provides a buffer that mitigates demand fluctuations, reducing the need for scaling actions.
Conversely, a lower over-provisioning rate requires more frequent scaling actions to maintain service assurance, leading to an increase in scaling actions.
Moreover, our closed-loop algorithm efficiently scales resources in response to traffic load changes without performance bottlenecks or outages, both in horizontal scaling (Fig. \ref{fig:s1_traffic_boundary}(b)) and vertical scaling (Fig. \ref{fig:s1_resources_utilization}). It accurately predicts future resource demands, enabling optimal resource allocation while maintaining service assurance. The algorithm also handles concurrent tasks across different network slices by deploying multiple {\PSA} instances, coordinated through the Broker and Slice Control components. This real-time coordination allows the algorithm to dynamically allocate resources to VNF instances across multiple network slices. For example, in response to a traffic spike on one network slice, the algorithm quickly allocates additional resources, preventing KPI violations. Simultaneously, when traffic decreases on another network slice, it reduces resource allocation, optimizing resource usage.


\section{Conclusion}
\label{sec:conclusion}

This paper presented {\PSA}, a proactive closed-loop algorithm for service assurance in 5G/B5G network slicing.
{\PSA} dynamically scales VNF resources and manages link capacity to meet network slice-specific KPIs while minimizing resource consumption.
By leveraging machine learning for traffic prediction and linear programming for resource optimization, {\PSA} proactively adapts to changing network conditions and prevents KPI violations.
Our experimental results demonstrate significant resource savings across diverse network slice types.
{\PSA} achieved up to 54.85\%, 50.87\%, 57.1\%, and 23.63\% resource savings for eMBB, mMTC, uRLLC, and VoIP slices, respectively, in comparison to the worst-case scenario.
Even with minimal over-provisioning at just 5\%, the {\PSA} algorithm remains highly effective, resulting in minimal KPI violations, with only 27 violations recorded over 24 hours of simulation.
These results highlight {\PSA}'s potential to significantly improve the efficiency and effectiveness of resource management in 5G/B5G networks.

In future studies, we intend to further explore and develop the capabilities of {\PSA} to manage dynamic network slice creation and deletion. The objective is to integrate more sophisticated traffic prediction models with the incorporation of network topology. Furthermore, the evaluation of {\PSA} is planned to be conducted in a real-world testbed.


\section*{Acknowledgment}
The first two authors of this paper received support for their internship from MITACS \& Ciena.

\bibliographystyle{IEEEtran}
\footnotesize
\bibliography{IEEEabrv,Biblio/references,Biblio/Close_Loop_SErvice_Assurance,Biblio/Online_learning,Biblio/5G_Slicing,Biblio/BJ,Biblio/ML,Biblio/Traffic_Prediction}

\vskip -2\baselineskip plus -1fil
\begin{IEEEbiography}[{\includegraphics[width=1in,height=1.25in,clip,keepaspectratio]{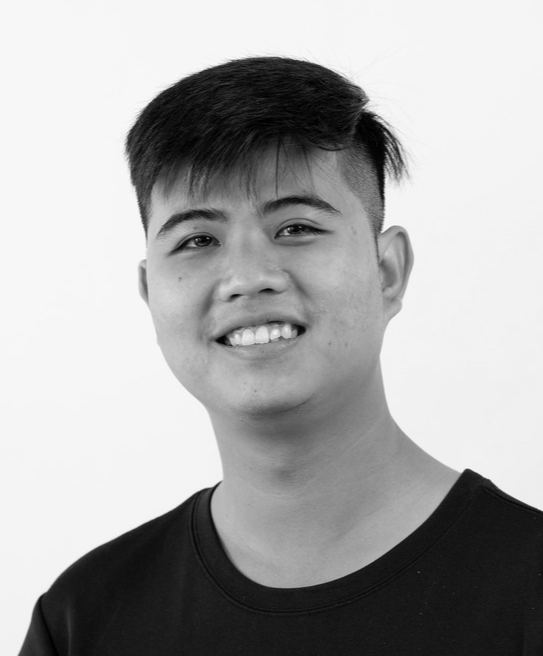}}]{Nguyen Phuc Tran} received his M.S. degree in Computer Science from the University of Information Technology, Vietnam National University, Ho Chi Minh City, in 2020. Since 2021, he has been pursuing his Ph.D. at Concordia University, Montreal, Quebec, Canada. With over five years of experience as a senior software engineer in system development and telecommunication technology, he has honed his expertise in system optimization, security, quality assurance, data analysis, team leadership, and stakeholder management. His current research interests encompass the design and application of artificial intelligence, including large language models, in mobile communication networks. He focuses on areas such as resource allocation, energy efficiency, green mobile networks, system design, root cause analysis, and system optimization.
\end{IEEEbiography}

\vskip -2\baselineskip plus -1fil
\begin{IEEEbiography}[{\includegraphics[width=1in,height=1.25in,clip,keepaspectratio]{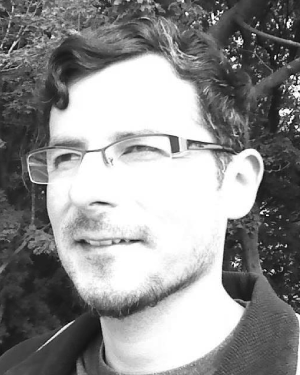}}]{Oscar Delgado} (Member, IEEE) received the M.A.Sc. degree from Concordia University, Montreal, QC, Canada, in 2010, and the Ph.D. degree in electrical engineering from McGill University, Montreal, in 2016. In 2017, he joined the Telecommunications and Signal Processing Laboratory (TSP), Department of Electrical and Computer Engineering, McGill University, where he is a Postdoctoral Researcher. His current research interests are in the applications of 5G wireless mobile communication technologies, including AI/machine learning, software-defined networks, network virtualization, and green wireless systems, and the analysis and design of video traffic management techniques, resource allocation strategies, and energy efficiency algorithms.
\end{IEEEbiography}

\vskip -2\baselineskip plus -1fil
\begin{IEEEbiography}[{\includegraphics[width=1in,height=1.25in,clip,keepaspectratio]{02_IEEE_TNSM/Fig_journal/brigitte.pdf}}]{Brigitte Jaumard}~(Senior Member, IEEE)~is the scientific director of Confiance IA, an Industrial research consortium - trustworthy AI supported by the Quebec government. She is also a professor in the Computer Science and Software Engineering (CSE) Department at Concordia University. Her research focuses on mathematical modelling and algorithm design (large-scale optimization and machine learning) for problems arising in communication networks, transportation and logistics networks. 
Recent studies include the design of efficient optimization/machine learning algorithms for network design, dimensioning and provisioning, scheduling in edge-computing and clouds, and 5G networks. During her 2020-2021 sabbatical year, she was a senior advisor for the Montreal Ericsson GAIA (Global Artificial Intelligence Accelerator) research center and the chief scientist of CRIM.

Brigitte Jaumard was ranked among the top 2\% of scientists in her field of research according to a 2021 study based on research citations. She was awarded several research chairs (Canada Research Chair and Concordia Research Chair, both Tier I during the years 2000-2019). B. Jaumard has published over 300 papers in international journals in Operations Research and in Telecommunications.

\end{IEEEbiography}

\end{document}